\documentclass[a4paper,11pt]{article}
\usepackage{a4wide}
\usepackage{float}
\usepackage{graphicx}
\usepackage[colorinlistoftodos]
{todonotes}
\usepackage[colorlinks=true, allcolors=purple]{hyperref}
\usepackage[super]{nth}
\usepackage{color}
\usepackage{amsmath}
\usepackage{amsthm}
\usepackage{amssymb}
\usepackage{caption}
\usepackage{subcaption}
\usepackage{xcolor}
\usepackage{soul}
\def\sech{\mathop{\rm sech}\nolimits}

\usepackage{comment}

\title{The Extended KdV Equation: Augmented Lagrangian and Variational Solitary Waves with Applications to Dispersive Hydrodynamics} 
\author{Saleh Baqer, Hamid Said\\
Department of Mathematics, College of Science, \\
Kuwait University, Sabah Al Salem University City, \\ P.O. Box 5969, Safat 13060, Shadadiya, Kuwait \and}
\date{}

\begin{document}

\maketitle

{\textit{Dedicated to N.F. Smyth and T.R. Marchant for their development of, and major contributions to, the extended KdV equation \cite{ekdv}}}

% Abstract

\begin{abstract}
In this work, we extend the method of averaged Lagrangian to the study of the general second-order (non-conservative) extended Korteweg--de Vries equation, known as the eKdV equation. Building on the framework introduced in \cite{hamidgk}, we construct a master (augmented) Lagrangian, modeled on Luke’s Lagrangian, that incorporates the governing constraints at the appropriate asymptotic orders via the method of Lagrange multipliers. Averaging the resulting Euler-Lagrange equations in the traveling wave setting yields the existence of a (single) solitary wave solution with a $\mathrm{sech}^{2}$ profile. Explicit second-order formulas are obtained for the height of the solitary wave, together with the solitary wave velocity and inverse width, in terms of a fixed amplitude parameter. A key feature of the derived expressions is their asymptotic reduction to the classical KdV results when the first-order terms are retained. To assess the robustness and utility of the variational solitonic solutions, the derived formulas are subsequently applied, via the dispersive shock equal amplitude approximation method, to estimate the height and velocity of the leading solitary wave edge of dispersive shock waves governed by the eKdV Riemann problem. Theoretical predictions for the relevant wave parameters in both the eKdV solitary wave and dispersive shock wave problems are compared with direct numerical simulations and found to be in strong agreement. 
\end{abstract}


\section{Introduction}
\label{sec:1}

The extended KdV equation (eKdV equation for short)
\begin{equation}\label{eKdV}
    u_{t} + u_x +\dfrac{3}{2} uu_{x} \alpha + \dfrac{1}{6}u_{3x} \beta +B_{1}u^2u_{x} \alpha^2 +B_{2}u_{5x} \beta^2 + \left( B_{3}u_{x}u_{xx}+B_{4}uu_{3x} \right) \alpha \beta =0,
\end{equation}
is a fifth-order nonlinear dispersive partial differential equation (PDE) arising in modeling a wide spectrum of dispersive hydrodynamic phenomena ranging from fluid mechanics \cite{ekdv,ekdhighermodu,annabook,annapaper,karimafifth,karimagardner,karimaphysicad,us,resekdv,salehekdv}, plasma physics \cite{ekdvplasma,salehaims}, nonlinear optics \cite{salehnem2}, and solid mechanics \cite{karimasolidbore1,karimasolidbore2,purohit}. From the prescriptive of modeling shallow water waves, this nonlinear dispersive wave equation can be obtained from the full set of Euler water wave equations (with an even bottom topography) under what is known as weakly nonlinear and shallow water (sometimes called long-wave) approximations. More precisely, within this model, after assuming that the wave displacement $u$ is invariant in the $y$-direction for instance so that $u=u(x,t)$, one carries a specified approximation scheme to second-order under the assumption that the displacement from the equilibrium surface is much smaller than the depth of the water container \cite{ekdv,annabook}.   In this case the coefficients take the values \cite{salehekdv,salehwwp}
\begin{equation} \label{coefs}
    B_1 = -3/8, \quad B_2= 19/360, \quad B_3= 23/24, \quad B_4 = 5/12,
\end{equation}
and the small non-dimensional parameters $\alpha$ and $\beta$ are related to the depth of the container. A thorough discussion on the derivation of the eKdV equation is given in Subsection \ref{subsec:1.1}. While the classical KdV equation (i.e., when $B_1 = B_2 = B_3 = B_4 = 0$) remains the simplest model for describing nonlinear long-wave propagation  with small amplitude, going beyond the first-order approximation is desirable in many situations arising in modeling various dispersive hydrodynamics. See for instance \cite{hamidgk} and references therein for detailed discussions on a wide range of applications.

Like its first-order approximation, the eKdV equation admits rich families of nonlinear wave solutions ranging from solitary waves (or solitons as coined for integrable systems \cite{ablowitzscattering}), rogue waves, nonlinear periodic cnoidal wavetrains \cite{ekdv,annabook,annapaper,karimafifth,karimagardner,karimaphysicad,tim_soliton,cnoidalekdv}. In fact, this higher order evolution amplitude equation arising in physical situations for specific choices of the coefficients in equation (\ref{eKdV}) have been studied extensively each on its own right, and each particular case can exhibit both distinct and complex behaviors. For instance, the \emph{Kawahara equation}, sometimes termed the fifth-order KdV equation, (when $B_1 = B_3 = B_4 =0$, but $B_2\neq 0$) produces monotone solitary wave solutions and other solitonic solutions with oscillatory tails (namely, resonant solitary wave propagation) \cite{resekdv,kawahara}, as well as classical and non-classical dispersive shock wave (DSW) solutions (\emph{alias}  undular bores as known in the context of fluids \cite{elreview}). This nonlinear dispersive wave equation appears when modeling capillary-gravity waves with none-negligible surface tension \cite{salehekdv,patkawahara,patkawtrav,pat,patjump}. Another nonlinear dispersive wave model is the \emph{Gardner equation} (when $B_{2}=B_3= B_4 =0$, and $B_1\neq 0$) which arises in different models such as plasmas \cite{gardnerplasma}; and the motion of internal, bright and dark, rogue waves in three layer fluids \cite{gardner}. On the other hand, the \emph{Gardner-Kawahar} equation (when $B_3= B_4 =0$, and $B_1, B_2 \neq 0$) arises in considering internal waves at the interface between two fluid layers; and also admits single and multi-solitary wave solutions with oscillatory asymptotics  \cite{karimafifth,gk2015}. Moreover, and recently \cite{hamidgk}, the authors considered a general form of the eKdV equation which conserves the corresponding energy (when $B_3 = 2B_4$ and $B_1, B_2 \neq 0$), termed the \emph{Conservative-eKdV equation}, and were successful in finding a single solitary wave solutions based on the method of averaged Lagrangian. Additionally, it was demonstrated that the equation inherits a Hamiltonian structure, and the derived solitonic solution was applied to the study of both classical and non-classical DSWs. It is important to note, however, that neither the eKdV equation \eqref{eKdV} nor the aforementioned evolution equations, except the Gardner equation, are integrable \cite{us}.   

The existence of an exact and closed-form solitary wave solution is not guaranteed for a broad class of dispersive hydrodynamic models, particularly those that lack integrability, including the eKdV equation. The availability of an explicit solitary wave solution—especially in the single $\sech^{2}$ form, due to its analytical tractability—has proven highly desirable, and often essential, for the study and analysis of various classical and non-classical dispersive hydrodynamic regimes. See \cite{salehekdv,salehnem2,salehwwp,equalamp,salehnem1} for a variety of relevant applications. Single solitary waves for the eKdV equation in the form of $u(x,t) = a_0\,\mathrm{sech}^2(w_s (x-V_st))$ (see Subsection \ref{subsec:3.1}) were first constructed by Karczewska {\em et al.} \cite{annabook,annapaper} for the specific problem of shallow water waves (i.e., when the coefficients are given by \eqref{coefs}). Using basic algebraic methods, the authors found an explicit expression for the speed of the solitary wave, which was determined a constant for all wave parameters $\alpha$ and $\beta$ unlike the classical KdV equation. Since the solitary wave speed belongs to (or embeds in) the continuous spectrum of the linear wave system, that is the regime of all possible linear phase velocities, the solitary wave solution found is an example of what is known as an (embedded) solitary wave. These types of solitary waves have known to exist in other contexts as well \cite{yang1,yang2}. A similar algebraic method was employed by \cite{karimafifth} to show the existence of single solitary solutions for equation \eqref{eKdV} for a particular set of coefficients, and the single solitary wave was constructed numerically for a multitude of special cases including the Gardner-Kawahara equation. Moreover, and more recently, Khusnutdinova \emph{et al.} \cite{karimaphysicad} implemented an asymptotic Kodama-Fokas-Liu near-identity transformation to map the eKdV equation to the Gardner equation whose soliton solutions are known and can therefore be used to estimate solitary wave solutions for the eKdV equation itself. 

The purpose of this paper is to find a (variational) single solitary wave solution to the eKdV equation \eqref{eKdV}. In this context, a single solitary wave refers to a solution of the eKdV equation with a $\sech^2$ profile. This profile is valid up to second-order in the wave parameters $\alpha$ and $\beta$ matching the same order of approximation of the equation of motion. We employ the variational approach developed by the authors in \cite{hamidgk} to derive this solution, and apply it to dispsersive hydrodynamics such as DSWs. This approach yields explicit expressions for the velocity, the amplitude and the (inverse) width of the solitary wave solution in terms of the coefficients $B_1, B_2, B_3$ and $B_4$ up to second-order in the parameters $\alpha$ and $\beta$. The classical results are then recovered if we restrict these expressions to first-order. These solutions are also compared with (direct) numerical solutions and applied to the study of classical DSWs in order to verify its efficacy and accuracy. In general, since the corresponding energy of the eKdV equation is not conserved (see equation \eqref{e:energycons} for example),  we do not expect that it would be amenable to direct Lagrangian methods (see \cite{hamidgk}).  However, as will be demonstrated, this problem becomes tractable once we consider a more general variational formulation, via what is known as \emph{Luke's Lagrangian}, distinct from what the authors had considered previously. 

In 1967 J.C. Luke formulated a variational principle to derive the shallow water waves equations \cite{whitham,lukevar}. Luke's Lagrangian involves the velocity potential and the free fluid surface, that is the wave displacement as unknowns. However, it was not until 1990 that Marchant and Smyth \cite{ekdv} applied  perturbation methods to obtain the eKdV equation (\ref{eKdV}) from Luke's variational principle. Since the eKdV equation appears in the second-order approximation, Marchant and Smyth carried out the perturbation in the corresponding Lagrangian up to third-order  in the relevant variables: the velocity potential and the wave displacement. While they were successful in obtaining a perturbed Lagrangian for the problem, it alone did not produce the eKdV equation. It needed to be supplemented with particular ansatz giving the velocity potential as a function of the wave displacement and its derivatives (up to second-order approximation), which once substituted into the Euler-Lagrange equations resulted in the eKdV equation \eqref{eKdV}. 

From a variational perspective, and interestingly enough, we find that the ansatz act as constraints on the system, and therefore can be accommodated in the overall analysis via the method of Lagrange multipliers. Once the \emph{augmented} Lagrangian is formulated, a variation of Whitham's averaged Lagrangian method is used to determine the wave parameters for the variational single solitary wave solution. We note that a careful examination of the derivation of the eKdV equation from the Euler shallow water system via the standard perturbation approach reveals the orders at which the constraints become relevant, a crucial fact in formulating the augmented Lagrangian. In this respect therefore, a brief discussion on the Euler shallow water wave equations and the derivation of the KdV and eKdV equations is warranted and presented next, and later scrutinized in Section \ref{sec:2}.  

\subsection{Shallow water waves, KdV and eKdV equations}\label{subsec:1.1}

Consider the motion of an incompressible, irrotational, and inviscid fluid under the influence of gravity where  the motion exhibits translational invariance with respect to the direction perpendicular to wave propagation (say the $y$-direction). Further assume a flat horizontal
bottom topography and a free surface given by the standard kinematic boundary condition. In non-dimensional form, and in terms of the velocity potential $\Phi (x,z,t)$ and the surface displacement $u(x,t)$ the equations of motion and the associated boundary conditions read \cite{annabook,whitham}

\begin{equation} \label{ww-intro}
  \left\{
  \begin{array}{llll}
  \vspace*{0.1 in}
\beta \Phi_{xx} + \Phi_{zz} = 0, \, \, \, \,  &\mathrm{for} \, \, \, \, 0 < z < 1+ \alpha u (x,t)    \\
\dfrac{1}{\beta} \Phi_z - \left( \alpha u_x \Phi_x + u_t \right) =0, \, \, \, \,  &\mathrm{for} \, \, \, \,  z = 1+ \alpha u (x,t)    \\
\Phi_t + \dfrac{1}{2} \left( \alpha \Phi_x^2 + \dfrac{\alpha}{\beta} \Phi_z^2  \right) + u =0, \, \, \, \,  &\mathrm{for} \, \, \, \,  z = 1+ \alpha u (x,t)    \\
\Phi_z = 0,   \, \, \, \,  &\mathrm{for} \, \, \, \,  z=0

    \end{array} 
    \right.
\end{equation}
where the non-dimensional quantities can be introduced as follows (the tildes are omitted after scaling):

\begin{equation} \label{non-dim-intro}
    \Tilde{\Phi} = \dfrac{h}{l a_0 \sqrt{gh}} \Phi, \quad \Tilde{x} = \dfrac{x}{l}, \quad \Tilde{u} = \dfrac{u}{A}, \quad \Tilde{z} = \dfrac{z}{H}, \quad \Tilde{t} = \dfrac{\sqrt{gh}}{l} t,
\end{equation}
with $\alpha = \dfrac{a_0}{h}$ and $\beta = \left( \dfrac{h}{l} \right)^2$; and the parameters $a_0, g, h,$ and $l$ denote the amplitude of the surface wave, the acceleration due to gravity in the $z$-direction, the depth of the container, and the mean wavelength, respectively. 

Next, assume that the parameters $\alpha,\beta<1$ (i.e., small amplitude and long wavelength approximations) and make the standard substitution
\begin{equation} \label{phi-1}
    \Phi(x,z,t) = \displaystyle\sum_{n=1}^\infty \phi_n(x,t) z^n,
\end{equation}
into equations \eqref{ww-intro}$_1$ and \eqref{ww-intro}$_4$ to obtain
\begin{equation} \label{phi-2}
\phi_n = \phi_{2m} = \dfrac{(-1)^m}{(2m)!} (\phi_0)_{2mx} \, \beta^m.
\end{equation}
Now by substituting this result into equations \eqref{ww-intro}$_2$ and \eqref{ww-intro}$_3$, and retaining terms up to $\mathcal{O}(\alpha, \beta)$ only one obtains the \emph{first-order Boussinesq equations}:
\begin{equation} \label{Bous-intro}
  \left\{
  \begin{array}{ll}
  \vspace*{0.1 in}
 u_x + w_t + \alpha w w_x - \dfrac{\beta}{2} w_{2xt} = 0, \\
 u_t + w_x + \alpha (u w)_x - \dfrac{\beta}{6} w_{3x} =0, \\
    \end{array} 
    \right.
\end{equation}
where we have denoted $w \doteq (\phi_0)_x$. The above system of equations can be further reduced into a single equation once the function $w$ is written in terms of the surface displacement $u$. One choice that ensures consistency up to $\mathcal{O}(\alpha, \beta)$ is 

\begin{equation} \label{anzats-intro}
    w = u - \dfrac{\alpha}{4} u^2 + \dfrac{\beta}{3} u_{xx}.
\end{equation}
Finally, we arrive at the classical KdV equation
\begin{equation} \label{KdV3-intro}
    u_t + u_x + \dfrac{3}{2} \alpha u u_x + \dfrac{1}{6} \beta u_{3x} = 0,
\end{equation}
where the wave parameters $\alpha$ and $\beta$ are of the same order\footnote{By setting $\alpha = \beta$, and the change of variables $x \to \sqrt{\dfrac{3}{2}} (x-t)$ and $t \to \dfrac{1}{4}\sqrt{\dfrac{3}{2}} \alpha t$, one obtains the KdV equation in standard form: $u_t + 6uu_x + u_{3x} =0$.}. See also the discussion leading to equation \eqref{Bous-ansatz}.
\bigskip

\textbf{Remark 1}. The traveling wave solutions of the classical KdV equation satisfy the equation 
\begin{equation} \label{KdV3-intro-theta}
    -V_{s}u_\theta + u_\theta + \dfrac{3}{2} \alpha u u_\theta + \dfrac{1}{6} \beta u_{3\theta} = 0,
\end{equation}
where $u(\theta) = u(x-V_{s}t)$ and $V_{s}$ is the speed of the traveling wave, which can be shown to satisfy the  expression \cite{whitham} 
\begin{equation} \label{V-intro}
    V_{s} = 1 +  \dfrac{a_0}{2}\alpha,
\end{equation}
where $a_0$ denotes the amplitude of the classical KdV soliton.

In many applications where the amplitude of the surface wave is no longer small, a natural step in extending the range of validity of the KdV model is to retain higher order terms in the approximation scheme sketched above. This is done by substituting equations \eqref{phi-1} and \eqref{phi-2} into the system \eqref{ww-intro} at the free surface; yielding the \emph{second-order} Boussinesq equations \cite{annabook}

\begin{equation} \label{Bous2-intro}
  \left\{
  \begin{array}{ll}
  \vspace*{0.1 in}
 u_x + w_t + \alpha w w_x - \dfrac{\beta}{2} w_{2xt} + \dfrac{\beta^2}{24} w_{4x t} + \dfrac{\alpha \beta}{2} \left( w_x w_{xx} -2 (u w_{xt})_x -w w_{3x} \right) = 0, \\
 u_t + w_x + \alpha (u w)_x - \dfrac{\beta}{6} w_{3x} + \dfrac{\beta^2}{120} w_{5x} -\dfrac{\alpha \beta}{2} (w_{xx} u)_x  =0. \\
    \end{array} 
    \right.
\end{equation}
In order to arrive at a single equation in the displacement field $u(x,t)$, Marchant and Smyth \cite{ekdv} extended the formula in \eqref{anzats-intro} to include second-order terms and found
\begin{equation} \label{anzats-2-intro}
      w = u - \dfrac{\alpha}{4} u^2 + \dfrac{\beta}{3} u_{xx} + \dfrac{\alpha^2}{8} u^3 + \dfrac{\beta^2}{10} u_{4x} + \alpha \beta \left( \dfrac{3}{16} u_x^2 + \dfrac{1}{2} u u_{xx}  \right).
\end{equation}
Substituting the previous equation into equation \eqref{Bous2-intro}$_2$ produces the eKdV equation for shallow water waves (see Subsection \ref{subsec:2.1}):
\begin{equation} \label{ekdv-intro}
     u_t + u_x + \dfrac{3}{2} \alpha u u_x + \dfrac{1}{6} \beta u_{3x} - \dfrac{3}{8} \alpha^2 u^2 u_x + \dfrac{19}{360} \beta^2 u_{5 x} + \alpha \beta \left( \dfrac{23}{24} u_x u_{xx}  + \dfrac{5}{12} u u_{3x} \right)  = 0.
\end{equation}
Clearly the above equation reduces to the classical KdV equation \eqref{KdV3-intro} when only terms that are linear in the parameters $\alpha$ and $\beta$ are retained. As will be shown in the next section, the ansatz \eqref{anzats-2-intro} must be supplemented by a constraint—given by equation \eqref{Bous-ansatz}—to ensure consistency in the Boussinesq system \eqref{Bous2-intro}. This observation will play a key role in deriving the variational formulation at the correct asymptotic orders.

The rest of the paper is organized as follows. In Section \ref{sec:2}, we review the main mathematical approach employed in this work; beginning with recalling Luke's variational  derivation of the Boussinesq, KdV and eKdV equations, and extending the authors' previous variational approach \cite{hamidgk} via the method of Lagrange multipliers leading to a general formulation of a master or augmented Lagrangian. These methods are then applied, in Section \ref{sec:3}, for constructing a variational single solitary wave solution to the eKdV equation \eqref{eKdV}. We begin by constructing a specific augmented Lagrangian for the problem where the ansatz are expressed as two distinct constraints. Then a variant of Whitham's averaged Lagrangian method is used to find the amplitude, width and velocity of the solitary wave under two different considerations.  The case of fixed amplitude parameter $a_{0}$ is first studied, and as a result explicit expressions for the velocity $V_{s}$ and the inverse width $w_{s}$ of the solitary wave are found up to second-order in the parameters $\alpha$ and $\beta$ in terms of this fixed amplitude parameter $a_{0}$. The classical results are recovered when we restrict these expressions to first-order in $\alpha$ and $\beta$. The second case concerns perturbations that may arise in the actual solitonic height $A_{s}$ itself as the higher order solitary wave propagates, for which there is strong numerical evidence (\cite{karimafifth,hamidgk}). A combination of energy and variatinoal methods is used to find an expression for the actual (higher order) solitonic height $A_{s}$ up to second-order (in $\alpha$ and $\beta$) in terms of the fixed amplitude parameter $a_{0}$. In Section \ref{sec:4}, these results are applied to the study of classical DSWs (shock solutions to the Riemann problem for the eKdV equation (\ref{eKdV})) via the use of the so-called DSW equal amplitude approximation method \cite{equalamp}, leading to formulas for the height and velocity of the leading solitary wave edge of the eKdV DSW. In Section \ref{sec:5}, our analytical results from Sections \ref{sec:3} and \ref{sec:4} are compared with direct numerical solutions of the eKdV equation (\ref{eKdV}). In addition, we discuss and compare our analytical and numerical results to previous works by Karczewska \emph{et al.} \cite{annabook,annapaper} on this subject, and finally Section \ref{sec:7} summarizes our findings.

\section{Mathematical approach}\label{sec:2}

In this section, we examine two topics key for a variational study of a single solitary wave solution to the eKdV equation. First, we examine the central role of Luke's Lagrangian \cite{whitham,lukevar} in the derivation of the second-order Boussinesq equations \eqref{Bous2-intro} and the eKdV equation for shallow water waves \eqref{ekdv-intro}. The derivation will serve in determining the necessary constraints that need to be taken into account when formulating the relevant Lagrangian. Second, we review the variational theory advanced by the authors' in their previous work \cite{hamidgk}, which will be further extended in the presence of point-wise constraints.

\subsection{Luke's Lagrangian and the eKdV equation}\label{subsec:2.1}

In the context of nonlinear water wave theory, the starting point for a variational derivation of the eKdV equation (\ref{eKdV}), as presented by Marchant and Smyth \cite{ekdv}, is Luke's Lagrangian, which in dimensionless variables reads (see also \cite{lukevar}): 
\begin{equation} \label{luke-lagrangian}
  L = \dfrac{1}{2} u^2 \alpha + \int_0^{1+\alpha u} \left( \Phi_t + \dfrac{1}{2} \alpha \Phi_x^2 + \dfrac{1}{2} \dfrac{\alpha}{\beta} \Phi_z^2 \right) dz,  
\end{equation}
where again $u = u(x,t)$ is the displacement of the wave,  and $\Phi = \Phi(x,z,t)$ is the velocity potential. We  substitute the expression for the velocity potential \eqref{phi-1} and \eqref{phi-2} into the above Lagrangian and retain terms up to $\mathcal{O}(\alpha^3, \alpha^2 \beta,  \alpha \beta^2)$, which gives
\begin{align} \label{lagranian(x,t)}
    L & = \phi_t + \alpha \left( u \phi_t + \dfrac{u^2}{2} + \dfrac{\phi_x^2}{2} \right) - \dfrac{1}{6} \beta \phi_{2xt} + \dfrac{1}{2} \alpha^2 u \phi_x^2 + \dfrac{1}{120} \beta^2 \phi_{4xt}   \nonumber \\
    & \mbox{} ~~~+ \alpha \beta \left( -\dfrac{1}{2} u \phi_{2xt} + \dfrac{1}{6} \phi_{xx}^2 - \dfrac{1}{6}\phi_x \phi_{3x} \right) + \alpha^2 \beta \left( -\dfrac{1}{2} u^2 \phi_{2xt} + \dfrac{1}{2} u \phi_{xx}^2 - \dfrac{1}{2} u \phi_x \phi_{3x}  \right) \nonumber \\
    & \mbox{} ~~~+ \alpha \beta^2 \left(  \dfrac{1}{24} u \phi_{4xt} + \dfrac{1}{40}  \phi_{3x}^2 - \dfrac{1}{30}  \phi_{xx} \phi_{4x} +\dfrac{1}{120}  \phi_{x} \phi_{5x} 
 \right),
\end{align}
where we have set $\phi \doteq \phi_0$, and hence $w = \phi_x$. Now the Euler-Lagrange equations read
\begin{equation} \label{EL-sec2}
  \left\{
  \begin{array}{lll}
  \vspace*{0.1 in}
\dfrac{\partial L}{\partial u} = 0, \vspace{0.07 in} \\
\dfrac{\delta L}{\delta \phi} = \dfrac{\partial L}{\partial \phi} - \dfrac{\partial }{\partial t} \left( \dfrac{\partial L}{\partial \phi_t} \right)  - \dfrac{\partial }{\partial x} \left( \dfrac{\partial L}{\partial \phi_x} \right) + \dfrac{\partial^2}{\partial x^2} \left( \dfrac{\partial L}{\partial \phi_{xx}} \right) - \dfrac{\partial^3}{\partial x^2 \partial t} \left( \dfrac{\partial L}{\partial \phi_{2xt}} \right) \vspace{0.03 in} \\
 - \dfrac{\partial^3}{\partial x^3} \left( \dfrac{\partial L}{\partial \phi_{3x}} \right) + \dfrac{\partial^4}{\partial x^4} \left( \dfrac{\partial L}{\partial \phi_{4x}} \right) - \dfrac{\partial^5}{\partial x^4 \partial t} \left( \dfrac{\partial L}{\partial \phi_{4xt}} \right) - \dfrac{\partial^5}{\partial x^5} \left( \dfrac{\partial L}{\partial \phi_{5x}} \right)  =  0. \\ \end{array}
    \right.
\end{equation}
Substituting for the Lagrangian \eqref{lagranian(x,t)} in the above system (\ref{EL-sec2}), yields (after dividing by the parameter $\alpha$)

\begin{equation} \label{EL-sec2-2}
  \left\{
  \begin{array}{lll}
  \vspace*{0.1 in}
\phi_t + u + \dfrac{\alpha}{2} \phi_x^2 - \dfrac{\beta}{2}\phi_{2xt} + \dfrac{\beta^2}{24} \phi_{4xt} + \alpha \beta \left( -u \phi_{2xt} + \dfrac{1}{2} \phi_{xx}^2 - \dfrac{1}{2} \phi_x \phi_{3x} \right) =0, \vspace{0.07 in} \\
u_t + \phi_{xx} + \alpha (u \phi_x)_x + \beta \left( - \dfrac{2}{3}\phi_{4x} - \dfrac{1}{2} u_{2xt} \right) + \beta^2 \left( \dfrac{2}{15}\phi_{6x} + \dfrac{1}{24} u_{4xt} \right) \vspace{0.005 in}\\
+ \alpha \beta \left( -\dfrac{1}{2} (u \phi_{3x})_x - (u \phi_{xx})_{xx} - (\dfrac{1}{2} u^2)_{2xt} - \dfrac{1}{2} (u \phi_x)_{3x} \right) = 0. \\
\end{array}
    \right.
\end{equation}
Clearly, we desire for the above Euler-Lagrange equations, which in this case is a system of two equations, to coincided with the second-order Boussinesq equations \eqref{Bous2-intro} introduced in Subsection \ref{subsec:1.1}. By differentiating equation \eqref{EL-sec2-2}$_1$ with respect to the $x$ variable, and setting $\phi_x = w$, we readily obtain the first equation in the Boussinesq system (i.e., equation \eqref{Bous2-intro}$_1$). It remains to show that equations \eqref{EL-sec2-2}$_2$ and \eqref{Bous2-intro}$_2$ coincide. In their current forms these two equations differ at the orders of  $\beta, \beta^2$ and $\alpha \beta$. 

A similar problem is addressed in \cite{whitham} in deriving the classical KdV equation \eqref{KdV3-intro} from the first-order Boussinesq equations \eqref{Bous-intro}. The expression for the velocity $w$ given in equation \eqref{anzats-intro} ensures  consistency of the system if the \emph{zeroth order} Boussinesq equation $u_t = -u_x + \mathcal{O}(\alpha , \beta)$ is imposed to exchange the time derivatives with the spatial ones at first-order in the parameter $\beta$. Hence, we expect that for deriving the eKdV equation \eqref{ekdv-intro} we need to impose the (first) order Boussinesq equation to facilitate the exchange of the derivatives at the correct orders. Specifically, we consider
\begin{equation} \label{Bous-ansatz}
    \phi_{xx} = - u_t - \alpha (\phi_x u)_x + \dfrac{\beta}{6} \phi_{4x}.
\end{equation}
Replacing the time derivative using the above expression at $\mathcal{O}(\beta, \beta^2, \alpha \beta)$ in \eqref{EL-sec2-2}$_2$, and retaining all terms of $\mathcal{O}(\beta^2, \alpha \beta)$  finally achieves equation \eqref{Bous2-intro}$_2$. In fact, a close look at the second-order system \eqref{Bous2-intro} reveals that any choice of $w$  ensuring consistency, such as equation \eqref{anzats-2-intro}, must also be accompanied by some rule allowing the interchange of spatial and temporal derivatives; this is precisely equation \eqref{Bous-ansatz}. 

We conclude from the brief discussion above that Luke's Lagrangian in the form \eqref{lagranian(x,t)} accompanied by the two equations \eqref{anzats-2-intro} and \eqref{Bous-ansatz} can produce the eKdV equation \eqref{ekdv-intro} (\emph{cf.} \cite{ekdv}).

\subsection{Variational theory}\label{subsec:2.2}

In this section, we briefly review the method of \emph{averaged Lagrangians} as formulated in \cite{hamidgk}. The Lagrange multipliers method will then be used to extend the previous results when multiple constraints are imposed over the problem.

Let us begin by considering  a functional of the form
\begin{equation} \label{m_lagrangian}
    \mathcal{L}(u) = \int_{-a}^a L(u, u_\theta, \cdots, u_{m\theta} ) \, d \theta . 
\end{equation}
Assume that the extremum $u = u(\theta) $ belongs to an appropriate admissible set and has the form:
\begin{equation} \label{anz-1}
u = u(z_1 (\theta), z_2(\theta), ..., z_n(\theta)),
\end{equation}
for some smooth vector function $ \Vec{z} = (z_1, \cdots ,z_n)$. Then the averaged Lagrangian in terms of these new variables reads 
\begin{equation} \label{av_lagrangian}
  \overline{\mathcal{L}}(\vec{z} ) =  \mathcal{L} \circ u(\vec{z}) = \int_{-a}^a L \left( u (\Vec{z}), u_\theta (\Vec{z}), \cdots, u_{m \theta}(\Vec{z}) \right) \, d \theta .
\end{equation}
Moreover, the Euler-Lagrange equations are satisfied in $\vec{z}$:
 \begin{equation} \label{EL0}
     \dfrac{\delta \overline{\mathcal{L}}}{ \delta \Vec{z}} = \vec{0}.
 \end{equation}
Or, equivalently 
\begin{equation} \label{EL}
\begin{array}{c} 
     \dfrac{\partial L}{\partial z_1} - \dfrac{d}{d \theta} \left( \dfrac{\partial L}{\partial z_1'} \right) + \dfrac{d^2}{d \theta^2} \left( \dfrac{\partial L}{\partial z_1''} \right) - \cdots + (-1)^m \dfrac{d^m}{d \theta^m} \left( \dfrac{\partial L}{\partial z_1^{(m)}} \right)   =  0, \\[4pt]
      \dfrac{\partial L}{\partial z_2} - \dfrac{d}{d \theta} \left( \dfrac{\partial L}{\partial z_2'} \right) + \dfrac{d^2}{d \theta^2} \left( \dfrac{\partial L}{\partial z_1''} \right) - \cdots + (-1)^m \dfrac{d^m}{d \theta^m} \left( \dfrac{\partial L}{\partial z_2^{(m)}} \right)   =  0, \\[4pt]
      \vdots \\[4pt]
      \dfrac{\partial L}{\partial z_n} - \dfrac{d}{d \theta} \left( \dfrac{\partial L}{\partial z_n'} \right) + \dfrac{d^2}{d \theta^2} \left( \dfrac{\partial L}{\partial z_1''} \right) - \cdots + (-1)^m \dfrac{d^m}{d \theta^m} \left( \dfrac{\partial L}{\partial z_n^{(m)}} \right)   =  0.
\end{array}
\end{equation}
for each $i = 1, 2, \cdots, n$. Now, we suppose that in addition to the field $u$ (and its derivative), the problem under consideration is dependent on another field $\phi$ (and its derivative) through a set of $k$ point-wise constraints of the form
\begin{equation}
\begin{array}{c}
G_1(u, \cdots, u_{m\theta}, \phi, \cdots ,\phi_{m\theta}) = 0, \\[4pt]
G_2(u, \cdots, u_{m\theta}, \phi, \cdots, \phi_{m\theta}) = 0, \\[4pt]
\vdots \\[4pt]
G_k(u, \cdots, u_{m\theta}, \phi, \cdots, \phi_{m\theta}) = 0.
\end{array}
\end{equation}
Then, the Lagrangian that is to be extremized in this case (which we term the \emph{augmented Lagrangian}) becomes
\begin{equation} \label{ave-aug-den}
L_{\mathrm{aug}}(u, u_\theta, \cdots, u_{m\theta}, \phi, \phi_\theta,  \cdots, \phi_{m\theta}) = L(u, u_\theta, \cdots, u_{m\theta}) + \displaystyle\sum_{i=1}^{k} \gamma_i G_i(u,  \cdots, u_{m\theta}, \phi, \cdots, \phi_{m\theta})
\end{equation}
where each Lagrange multiplier may be $\theta$ dependent, that is $\gamma_i = \gamma_i(\theta)$, for every $i=1,2, \cdots k$. Therefore, under the ansatz 
\begin{equation} \label{anz-2}
u = u(z_1 (\theta), z_2(\theta), ..., z_n(\theta)), \qquad \phi = \phi(z_1 (\theta), z_2(\theta), ..., z_n(\theta)),
\end{equation} 
the Euler-Lagrange equations \eqref{EL} are satisfied with  $L$ replaced by $L_{\mathrm{aug}}$.
\bigskip

\textbf{Remark 2}. Throughout this paper as was done in \cite{hamidgk}, all fields are taken, in addition, to have  support over $(-a,a)$ or to decay sufficiently fast enough in the limit $|x| \to \infty$ if the spatial domain is taken to be $\mathbb{R}$.

\section{Variational solitary waves}\label{sec:3}

We are now in position to detail our variational study of single solitary wave solutions to the eKdV equation \eqref{eKdV}. As such, we begin by defining the phase $\theta = x-V_{s}t$, where $V_{s}$ denotes the travelling wave velocity. The Lagrangian \eqref{lagranian(x,t)} in the phase $\theta$ becomes
\begin{align} \label{lagranian(theta)}
    L & = -V_s\phi_\theta  + \alpha \left( -V_s u \phi_\theta + \dfrac{u^2}{2} + \dfrac{\phi_\theta^2}{2} \right) + \dfrac{V_s}{6} \beta \phi_{3\theta} + \dfrac{1}{2} \alpha^2 u \phi_\theta^2 - \dfrac{V_s}{120} \beta^2 \phi_{5\theta} \nonumber \\ 
    & \mbox{} ~~~ + \alpha \beta \left( \dfrac{V_s}{2} u \phi_{3\theta} + \dfrac{1}{6} \phi_{\theta\theta}^2 - \dfrac{1}{6}\phi_\theta \phi_{3\theta} \right) + \alpha^2 \beta \left( \dfrac{V_s}{2} u^2 \phi_{3\theta} + \dfrac{1}{2} u \phi_{\theta\theta}^2 - \dfrac{1}{2} u \phi_\theta \phi_{3\theta}  \right) \nonumber \\
    & \mbox{} ~~~ + \alpha \beta^2 \left(  \dfrac{-V_s}{24} u \phi_{5\theta} + \dfrac{1}{40}  \phi_{3\theta}^2 - \dfrac{1}{30} u \phi_{\theta \theta} \phi_{4\theta} +\dfrac{1}{120} u \phi_{\theta} \phi_{5 \theta} 
 \right).
\end{align}
As mentioned previously, the Lagrangian \eqref{lagranian(x,t)} needs to be supplemented with specific  ansatzes, related to the two fields $u$ and $\phi$, to be able to produce the eKdV equation \eqref{eKdV}. By requiring equation \eqref{anzats-2-intro} to be satisfied, we write the following constraint in terms of the phase $\theta$:
\begin{equation} \label{G(theta)}
    G(u, ...,u_{4\theta}, \phi_\theta) = \phi_\theta - \left( u - \dfrac{1}{4} \alpha u^2 + \frac{1}{3} \beta u_{\theta \theta} + C_1 \alpha^2 u^3 +C_2 \beta^2 u_{4 \theta} + \alpha \beta \left( C_3 u_\theta^2 + C_4 u u_{\theta \theta} \right) \right) = 0.
\end{equation}
General coefficients $C_1, C_2, C_3$, and $C_4$ are chosen for the constraint so to adapt to various physical settings; that is they are related to the choices of coefficients $B_1, B_2, B_3$, and $B_4$. See Appendix \ref{appa} for explicit relations between these coefficients; in fact it is shown that the eKdV equation (in terms of phase $\theta$) can be obtained without any reference to the coefficient $C_3$. Henceforth we assume that $C_3\equiv 0$. 

As we have already mentioned in Section \ref{sec:2}, the Lagrangian \eqref{lagranian(theta)} together with constraint \eqref{G(theta)} alone cannot reproduce the eKdV equation. Therefore, careful inspection of a suitable constraint in the form of equation \eqref{Bous-ansatz} is required. Firstly, we reiterate that the orders at which the corrections in the Euler-Lagrange equations, particularly equation \eqref{EL-sec2-2}$_2$, are necessary are $\beta, \beta^2$, and $\alpha \beta$. However, the pertinent terms in the constraint \eqref{Bous-ansatz} only appear at the orders: zero,  $\alpha$ and $ \beta$. Another associated consideration is related to the use of the averaging method (see Subsection \ref{subsec:3.1}). The application of the method entails having a Lagrangian in terms of  one field only. For our problem, it is natural to make this choice for $u$ (and its derivatives), implying that $\phi_\theta$  needs to be expressed in terms of $u$. This is in fact the case in the constraint \eqref{G(theta)} in contrast to \eqref{Bous-ansatz}. The latter consideration is dealt with by substituting equation \eqref{G(theta)} into $\eqref{Bous-ansatz}$ and retaining terms up to $\mathcal{O}(\alpha, \beta)$:
\begin{equation} \label{2nd-const}
    (u - V_s u) +\dfrac{3 \alpha}{4} u^2 + \dfrac{\beta}{6} u_{\theta \theta} = 0.
\end{equation}
 However, we are still left with the issue of the discrepancies in the orders of magnitude. To remedy the situation, we multiply the previous equation by $\alpha \beta$ in order to correct the Euler-Lagrange equations at $\mathcal{O}(\beta, \beta^2, \alpha \beta)$. Hence, we define the constraint
\begin{equation} \label{H(theta)}
  H(u, u_{\theta \theta}) =  (u - V_s u) \alpha \beta +\dfrac{3 \alpha^2 \beta}{4} u^2 + \dfrac{\alpha \beta^2}{6} u_{\theta \theta} = 0,
\end{equation}
where it is expected that, as in the calculations undertaken in Section \ref{sec:2} a division by $\alpha$ will ensue in the final equations.  

The \emph{augmented Lagrangian} for our problem, finally, becomes
\begin{equation} \label{aug-L(theta)}
  L_{\rm aug} = L + \lambda G + \sigma H,
\end{equation}
where $\lambda$ and $\sigma$ denote two Lagrange multipliers associated with the constraints. We conclude this subsection by making two remarks. First, since the problem, as formulated in terms of the above Lagrangian $L_{\mathrm{aug}}$, depends on the wave velocity $V_s$, it is expected that the coefficients $C_1, C_2$, and $C_4$ that need to be determined, and the Lagrange multipliers $\lambda$ and $\sigma$, all depend on $V_s$ as well. This is illustrated for the case of the water wave problem in Appendix \ref{appa}. Second, as the classical KdV equation \eqref{KdV3-intro} demonstrates, the velocity $V_s$ can depend on the parameters $\alpha$ and $\beta$. As we are considering the eKdV equation \eqref{eKdV}, one expects that the dependence on $\alpha$ and $\beta$ should go up to second-order, namely, 
\begin{equation} \label{amp-sec-order}
    V_s  = 1 + \alpha \dfrac{a_0}{2} + \mathcal{O}(\alpha^2, \beta^2, \alpha \beta).
\end{equation}
An explicit formula will be found later in this section. 

\subsection{Variational method} \label{subsec:3.1}

We start by implementing the averaged Lagrangian method \cite{hamidgk}. We seek a solitary wave solution of the form
\begin{equation} \label{soliton_Riesz_general}  
u (\theta) =a_{0}\mathrm{sech}^2(w_{s}\,\theta),
\end{equation}
where $w_{s}$ is the inverse width of the solitary wave, and $a_{0}$ is a (fixed) amplitude parameter. Our goal is to find an expression for $w_s$ and $V_s$ in terms of $a_0$ up to second-order in the wave parameters $\alpha$ and $\beta$. We shall further assume that the derivatives of $a_0$ and $w_s$ are neglected as being small. An expression for $\phi_\theta$ can be found in terms of the wave parameters due to the constraint \eqref{G(theta)} and then substituted into the Lagrangian \eqref{lagranian(theta)} before taking averages. This is equivalent to expressing the augmented Lagrangian \eqref{aug-L(theta)} in terms of $u$ (and its derivatives) alone with the assistance of the constraint \eqref{G(theta)}. This is done next.

\subsubsection{An equivalent Lagrangian}\label{subsubsec:3.1.1}

We begin by substituting the constraint \eqref{G(theta)} into the Lagrangian $L$ in equation \eqref{lagranian(theta)}, and retaining terms up to $\mathcal{O}(\alpha^3, \alpha^2 \beta, \alpha \beta^2 )$
\begin{align}\label{aug-L-av}
    L & = - V_su   +\left( \dfrac{-3 V_s}{4} u^2 + u^2  \right) \alpha - \left( \dfrac{V_s}{6} u_{\theta \theta} \right) \beta  + \left( \dfrac{V_s}{4} u^3 + \dfrac{1}{4} u^3 - V_s C_1 u^3 \right) \alpha^2 \nonumber \\
    & \mbox{} ~~~ +\left( \dfrac{17 V_s}{360}  u_{4\theta} -V_sC_2 u_{4 \theta} \right) \beta^2 + \Bigg( \dfrac{1}{6} u u_{\theta \theta} - \dfrac{V_s}{12} u_\theta^2 + \dfrac{1}{6} u_\theta^2  + \frac{V_s}{12} u u_{\theta \theta}  - V_s C_4 u u_{\theta \theta} \Bigg) \alpha \beta \nonumber \\
    & \mbox{} ~~~ + \left(\frac{5}{12} u u_\theta^2 -\frac{1}{8} u^2 u_{\theta \theta} +  +C_4 u^2 u_{\theta \theta}  -\frac{V_s}{4} u u_{\theta}^2 + \frac{V_s}{4} u^2 u_{\theta \theta}    -  V_s C_4 u^2 u_{\theta \theta} + V_s C_1 u u_\theta ^2 \right. \nonumber \\ 
     & \mbox{} ~~~ \left. + \frac{V_s}{2} C_1 u^2 u_{\theta \theta} \right)  \alpha^2 \beta + \left( \dfrac{7}{90} u_{\theta} u_{3\theta} - \dfrac{17}{360} u u_{3\theta} + \dfrac{V_s}{80} u_{\theta \theta}^2 + \dfrac{1}{40} u_{\theta \theta}^2 + \dfrac{V_s}{60}  u_{\theta} u_{3\theta} + \dfrac{31V_s }{240} u u_{4\theta}   \right. \nonumber \\
    & \mbox{} ~~~ \left. + \dfrac{V_s}{6} C_4 u_{\theta \theta}^2 - V_s C_2 u u_{4 \theta}  + \dfrac{V_s}{3} C_4 u_\theta u_{3 \theta} + \dfrac{V_s}{6} C_4 u u_{4 \theta} + C_2 u u_{4\theta} \right) \alpha \beta^{2}  \nonumber \\
    & \mbox{} ~~~ + \left( C_1 u^4 -\dfrac{7}{32} u^4 -V_s  C_1u^4  \right) \alpha^3. 
\end{align}
The corresponding equivalent augmented Lagrangian
\begin{equation} \label{aug-L-hat}
    L_{\mathrm{aug}} = L + \lambda G + \sigma H,
\end{equation}
is now a function of the phase $\theta$, the field $u$ and its derivatives, and the velocity potential $\phi_\theta$. Here, the functions $G$ and $H$ represents the constraints \eqref{G(theta)} and \eqref{H(theta)}, respectively. As was demonstrated in Appendix \ref{appa}, the multiplier $\lambda$ is constant so the term $\lambda G$ can be readily averaged. However, we found that $\sigma = \frac{4}{3}(2B_4 - B_3) u_{\theta \theta}$; hence $L_{\rm aug}$ cannot be averaged in order to get a bona fide averaged Lagrangian. We overcome this challenge using the results of Subsection \ref{subsec:2.2}.

%+ \lambda \left( \phi_\theta - \eta + \dfrac{1}{4} \eta^2 \alpha - \dfrac{1}{3} \eta_{\theta \theta } \beta - C_1 \eta^3 \alpha^2 - C_2 \eta_{4\theta} \beta^2 - (C_3 \eta_\theta^2 + C_4 \eta \eta_{\theta \theta}) \alpha \beta  \right)

%\begin{align*} \label{aug-L-av}
  %  \overline{\mathcal{L}}_{\mathrm{aug}} (a_0, w_s)= \int_{-\infty}^\infty L_{\mathrm{aug }| u = u(\theta)}  \, d\theta = \left ( \dfrac{-2 V_s a_0}{w_s} - \dfrac{2 a_0 \lambda}{w_s} \right) + \left( \dfrac{\lambda a_0^2}{3w_s} -\dfrac{V_s a_0^2}{w_s} + \dfrac{4 a_0^2}{3w_s} \right) \alpha \\
   %+ \Bigg( \dfrac{-16V_s}{15}C_3 a_0^2 w_s + \dfrac{16V_s}{15}C_4 a_0^2 w_s - \dfrac{16 \lambda }{15}C_3 a_0^2 w_s  + \dfrac{16 \lambda}{15}C_4 a_0^2 w_s 
   %- \dfrac{8 V_s}{45} a_0^2 w_s \Bigg) \alpha \beta \\
    %+ \Bigg( \dfrac{4 a_0^3}{15 w_s} - \dfrac{16a_0^3 V_s C_1}{15 w_s} - \dfrac{16a_0^3 \lambda C_1}{15 w_s}  + \dfrac{4a_0^3 V_s}{15 w_s}\Bigg) \alpha^2 \\
    %+ \Bigg( \dfrac{128}{315} a_0^3 w_s - \dfrac{64 V_s}{105} a_0^3 C_3 w_s + \dfrac{128 V_s}{105} a_0^3 C_4 w_s -\dfrac{16 V_s}{35} a_0^3 w_s \\
    %+ \dfrac{64}{105} a_0^3 C_3 w_s - \dfrac{128}{105} a_0^3 C_4 w_s\Bigg)  \alpha^2 \beta  \\
   %+ \Bigg( \dfrac{-32}{105} a_0^2 w_s^3 - \dfrac{64 V_s}{21} a_0^2 C_2 w_s^3 + \dfrac{8 V_s}{21} a_0^2 w_s^3 + \dfrac{64 }{21} a_0^2 C_2 w_s^3\Bigg) \alpha \beta^2 \\
   %+ \Bigg( \dfrac{-1}{5 w_s} a_0^4 - \dfrac{32 V_s} {35 w_s} C_1 a_0^4  + \dfrac{32 }{35 w_s} C_1 a_0^4 \Bigg) \alpha^3 \numberthis
%\end{align*}

\subsubsection{Averaging method}\label{subsubsec:3.1.2}
The parameters $a_0$ and $w_s$ must satisfy an Euler-Lagrange system resulting from the Lagrangian \eqref{aug-L-hat} in accordance to equation \eqref{EL}
\begin{equation} \label{EL_av_3}
  \left\{
  \begin{array}{ll}
  \vspace*{0.1 in}
  \dfrac{\partial L}{\partial a_0 } + \lambda \dfrac{\partial G}{\partial a_0 } + \sigma(\theta) \dfrac{\partial H}{\partial a_0 } = 0, \\
\dfrac{\partial L}{\partial w_s } + \lambda \dfrac{\partial G}{\partial w_s } + \sigma(\theta) \dfrac{\partial H}{\partial w_s } = 0.
    \end{array} 
    \right.
\end{equation}
In these equations, the quantity $\sigma$ can be expressed in terms of $a_0, w_s$ and $\theta$ by employing the solitary wave solution \eqref{soliton_Riesz_general}. This substitution then allows the Euler–Lagrange equations to be averaged over the phase $\theta \in (-\infty, \infty) $, resulting in a system depending only on $a_0, w_s, V_s, \lambda$, together with the parameters $\alpha$ and $\beta$. Moreover, as noted in Remark 2 of Subsection 2.2, the averaging procedure eliminates the dependence on the velocity potential $\phi_\theta$ that appears in the constraint defining $G$.

The Euler-Lagrange equation corresponding to taking variation with respect to $a_0$ (i.e. equation \eqref{EL_av_3}$_1$ after averaging) reads 
\begin{eqnarray}\label{EL-a}
    & & \left (-2 \dfrac{ V_s}{w_s} - 2 \dfrac{ \lambda}{w_s} \right) + \left( \dfrac{2}{3}\dfrac{ \lambda a_0}{w_s} -2 \dfrac{ V_s a_0}{w_s} + \dfrac{8}{3} \dfrac{a_0}{w_s} \right) \alpha + \left( \dfrac{64}{45} w_s B_3 a_0(1-V) +\dfrac{32 }{15} V_s C_4 a_0 w_s  \right. \nonumber \\
    & & \mbox{} \left. + \dfrac{128}{45} w_s B_4 a_0(V-1)
   + \dfrac{32 }{15} \lambda C_4 a_0 w_s - \dfrac{16 }{45} V_s a_0 w_s \right)\alpha\beta + \left( \dfrac{4}{5} \dfrac{a_0^2}{w_s} - \dfrac{16}{5} \dfrac{a_0^2 V_s C_1}{ w_s}  - \dfrac{16}{5} \dfrac{a_0^2 \lambda C_1}{ w_s}  \right. \nonumber \\
   & & \mbox{} \left. + \dfrac{4}{5}\dfrac{a_0^2 V_s}{ w_s} \right)\alpha^{2} + \left( \frac{256}{105}B_3 a_0^2 w_s - \dfrac{512}{105} B_4 a_0^2 w_s  + \dfrac{128}{105} a_0^2 w_s  + \dfrac{128}{35} V_s a_0^2 C_4 w_s -\dfrac{48 }{35}V_s a_0^2 w_s   \right. \nonumber \\
   & & \mbox{} \left.  - \dfrac{128}{35} a_0^2 C_4 w_s \right)\alpha^{2}\beta + \left( \dfrac{256}{189}B_4 w_s^3 a_0 - \dfrac{128}{189} B_3 w_s^3 a_0 -\dfrac{64}{105} a_0 w_s^3 - \dfrac{128 }{21} V_s a_0 C_2 w_s^3  \right. \nonumber \\
   & & \mbox{} \left. + \dfrac{16 }{21} V_s a_0 w_s^3 + \dfrac{128}{21} a_0 C_2 w_s^3 \right)\alpha\beta^{2} +\left( \frac{128}{35} \dfrac{ C_1 a_0^3 }{ w_s}    - \frac{128}{35} \dfrac{ V_s C_1 a_0^3} { w_s} -\dfrac{4}{5}\dfrac{a_0^3}{ w_s}  \right)\alpha^{3}=0,
\end{eqnarray}
\begin{comment}
\begin{align*} \label{EL-a}
    \left (-2 \dfrac{ V_s}{w_s} - 2 \dfrac{ \lambda}{w_s} \right) + \left( \dfrac{2}{3}\dfrac{ \lambda a_0}{w_s} -2 \dfrac{ V_s a_0}{w_s} + \dfrac{8}{3} \dfrac{a_0}{w_s} \right) \alpha \\
   + \Bigg( \dfrac{64}{45} w_s B_3 a_0(1-V)+ \dfrac{128}{45} w_s B_4 a_0(V-1)  +\dfrac{32 }{15} V_s C_4 a_0 w_s + \dfrac{32 }{15} \lambda C_4 a_0 w_s 
   - \dfrac{16 }{45} V_s a_0 w_s \Bigg) \alpha \beta \\
    + \Bigg( \dfrac{4}{5} \dfrac{a_0^2}{w_s} - \dfrac{16}{5} \dfrac{a_0^2 V_s C_1}{ w_s} - \dfrac{16}{5} \dfrac{a_0^2 \lambda C_1}{ w_s}  + \dfrac{4}{5}\dfrac{a_0^2 V_s}{ w_s}\Bigg) \alpha^2 \\
    + \Bigg( \frac{256}{105}B_3 a_0^2 w_s - \dfrac{512}{105} B_4 a_0^2 w_s  + \dfrac{128}{105} a_0^2 w_s  + \dfrac{128}{35} V_s a_0^2 C_4 w_s -\dfrac{48 }{35}V_s a_0^2 w_s  - \dfrac{128}{35} a_0^2 C_4 w_s\Bigg)  \alpha^2 \beta  \\
   + \Bigg( \dfrac{256}{189}B_4 w_s^3 a_0 - \dfrac{128}{189} B_3 w_s^3 a_0 -\dfrac{64}{105} a_0 w_s^3 - \dfrac{128 }{21} V_s a_0 C_2 w_s^3 + \dfrac{16 }{21} V_s a_0 w_s^3 + \dfrac{128}{21} a_0 C_2 w_s^3 \Bigg) \alpha \beta^2 \\
   + \Bigg(\frac{128}{35} \dfrac{ C_1 a_0^3 }{ w_s}    - \frac{128}{35} \dfrac{ V_s C_1 a_0^3} { w_s} -\dfrac{4}{5}\dfrac{a_0^3}{ w_s} \Bigg) \alpha^3 = 0 \numberthis
\end{align*}
\end{comment}
while the Euler-Lagrange equation resulting from  taking variation with respect to $w_s$ is
\begin{eqnarray}\label{EL-w}
    & & \left ( 2 \dfrac{ V_s a_0}{w_s^2} + 2\dfrac{ a_0 \lambda}{w_s^2} \right) + \left( -\dfrac{1}{3} \dfrac{\lambda a_0^2}{w_s^2} +\dfrac{V_s a_0^2}{w_s^2} - \dfrac{4}{3} \dfrac{ a_0^2}{w_s^2} \right)  \alpha + \left( \dfrac{32}{45} a_0^2 (1-V_s) (B_3-2B_4) - \dfrac{8 }{45} V_s a_0^2  \right. \nonumber \\
    & & \mbox{} \left. + \dfrac{16 }{15} \lambda C_4 a_0^2 + \dfrac{16}{15 } V_s C_4 a_0^2  \right)\alpha\beta +\left( -\dfrac{4}{15} \dfrac{ a_0^3}{ w_s^2} + \dfrac{16}{15} \dfrac{a_0^3 V_s C_1}{ w^2_s} + \dfrac{16}{15}\dfrac{a_0^3 \lambda C_1}{ w^2_s}  - \dfrac{4}{15} \dfrac{a_0^3 V_s}{ w^2_s} \right)\alpha^{2} \nonumber \\
     & & \mbox{} +\left( \dfrac{32}{315} a_0^3 (B_3 - 2B_4)  + \dfrac{128 V_s}{105} a_0^3 C_4 -\dfrac{16 V_s}{35} a_0^3 + \dfrac{128}{315} a_0^3 \right)\alpha^{2}\beta +\left( \dfrac{64}{63}a_0^2w^2(2B_4-B_3)   \right. \nonumber \\
      & & \mbox{} \left. - \dfrac{32}{35} a_0^2 w_s^2 - \dfrac{64 }{7} V_s a_0^2 C_2 w_s^2 + \dfrac{8 }{7} V_s a_0^2 w_s^2 + \dfrac{64 }{7} a_0^2 C_2 w_s^2 \right)\alpha\beta^{2} + \left( \dfrac{1}{5 } \dfrac{a_0^4}{w^2_s} + \dfrac{32}{35} \dfrac{C_1 a_0^4  V_s} { w^2_s} \right. \nonumber \\ 
      & & \mbox{} \left. - \dfrac{32 }{35 } \dfrac{C_1 a_0^4}{w^2_s} \right)\alpha^{3}=0.
\end{eqnarray}
\begin{comment}
\begin{align*} \label{EL-w}
    \left ( 2 \dfrac{ V_s a_0}{w_s^2} + 2\dfrac{ a_0 \lambda}{w_s^2} \right) + \left( -\dfrac{1}{3} \dfrac{\lambda a_0^2}{w_s^2} +\dfrac{V_s a_0^2}{w_s^2} - \dfrac{4}{3} \dfrac{ a_0^2}{w_s^2} \right) \alpha \\
   + \Bigg( \dfrac{32}{45} a_0^2 (1-V_s) (B_3-2B_4) + \dfrac{16}{15 } V_s C_4 a_0^2 + \dfrac{16 }{15} \lambda C_4 a_0^2 - \dfrac{8 }{45} V_s a_0^2 \Bigg) \alpha \beta \\
    + \Bigg( -\dfrac{4}{15} \dfrac{ a_0^3}{ w_s^2} + \dfrac{16}{15} \dfrac{a_0^3 V_s C_1}{ w^2_s} + \dfrac{16}{15}\dfrac{a_0^3 \lambda C_1}{ w^2_s}  - \dfrac{4}{15} \dfrac{a_0^3 V_s}{ w^2_s}\Bigg) \alpha^2 \\
    + \Bigg( \dfrac{32}{315} a_0^3 (B_3 - 2B_4)  + \dfrac{128 V_s}{105} a_0^3 C_4 -\dfrac{16 V_s}{35} a_0^3  
     + \dfrac{128}{315} a_0^3 \Bigg)  \alpha^2 \beta  \\
   + \Bigg(\dfrac{64}{63}a_0^2w^2(2B_4-B_3)  - \dfrac{32}{35} a_0^2 w_s^2 - \dfrac{64 }{7} V_s a_0^2 C_2 w_s^2 + \dfrac{8 }{7} V_s a_0^2 w_s^2 + \dfrac{64 }{7} a_0^2 C_2 w_s^2\Bigg) \alpha \beta^2 \\
   + \Bigg( \dfrac{1}{5 } \dfrac{a_0^4}{w^2_s} + \dfrac{32}{35} \dfrac{C_1 a_0^4  V_s} { w^2_s}  - \dfrac{32 }{35 } \dfrac{C_1 a_0^4}{w^2_s} \Bigg) \alpha^3 = 0
   \end{align*}
\end{comment}
Proceeding as in Appendix \ref{appa}, and expressing $C_1, C_2,$ and $C_4$ in terms of $B_1, B_2, B_3,$ and $B_4$ (see equation \eqref{a-sys-sol}), equation \eqref{EL-w} can be solved for the velocity $V_s$ in terms of the width $w_s$ and the amplitude parameter $a_0$, yielding
\begin{align}\label{V_s_w_s}
    V_s & = 1 + \dfrac{2}{5} a_0 \alpha + \dfrac{2}{15} w_s^2 \beta + \dfrac{4}{35} B_1 a_0^2 \alpha^2  + \left( -\dfrac{32}{105} B_4 a_0 w_s^2 + \dfrac{8}{21} B_3 a_0 w_s^2 \right) \alpha \beta \nonumber \\
     & \mbox{} ~~~ + \left(- \dfrac{48}{7} B_2 w_s^4 + \dfrac{16}{21}B_3 w_s^4 -\dfrac{32}{21} B_4 w_s^4  \right)\beta^2.
\end{align}
\begin{comment}
\begin{align*} \label{V_s_w_s}
       V_s = 1 + \dfrac{2}{5} a_0 \alpha + \dfrac{2}{15} w_s^2 \beta + \dfrac{4}{35} B_1 a_0^2 \alpha^2  + \left( -\dfrac{32}{105} B_4 a_0 w_s^2 + \dfrac{8}{21} B_3 a_0 w_s^2 \right) \alpha \beta \\
       + \left(- \dfrac{48}{7} B_2 w_s^4 + \dfrac{16}{21}B_3 w_s^4 -\dfrac{32}{21} B_4 w_s^4  \right)\beta^2 \numberthis
   \end{align*}
\end{comment}
Substituting the above expression into the Euler-Lagrange equation \eqref{EL-a}, and retaining terms up to second-order in $\alpha$ and $\beta$ gives a quartic polynomial in the unknown parameter $w_s$:
\begin{multline} \label{quintic}
 \left( \dfrac{64}{21} a_0 \beta^2  B_2 - \dfrac{64}{189} a_0 \beta^2  B_3 + \dfrac{16}{189} a_0 \beta^2 B_4 \right) w_s^4 +
    \left(- \dfrac{16}{315} a_0^2 \alpha \beta B_3 - \dfrac{88}{315} a_0^2 \alpha \beta B_4 - \dfrac{4 }{45} a_0 \beta   \right)w_s^2 \\
    + \left(\dfrac{4}{105}  a_0^3 \alpha^2 B_1 +\dfrac{1}{15} a_0^2 \alpha \right) = 0.
\end{multline}
The four roots of this polynomial are
\begin{equation} \label{ws_roots}
  w_s = \pm \dfrac{\sqrt{30}}{40 \sqrt{\beta K_1}} \sqrt{\alpha K_2 \mp \sqrt{K_3 \alpha^2+K_4 \alpha + 49} + 7},
\end{equation}
where the coefficients
\begin{align}
    &K_1 = 9B_2 - B_3 + 2B_4, \\
    &K_2 = a_0( 4B_3 + 22 B_4), \\
    &K_3 = 8 a_0^2 \left( -360 B_1 B_2 +40 B_1 B_3 - 80 B_1B_4 + 2 B_3^2 + 22 B_3 B_4 + \dfrac{121}{2} B_4^2 \right), \\
   & K_4 =  28 a_0\left(-180 B_2 + 22 B_3 -29 B_4  \right) .
\end{align}
In order to find the correct width, we need to expand the roots in powers of $\alpha$ and $\beta$ and verify which gives the correct result at $\mathcal{O}(1)$, namely
\begin{equation}\label{w0}
    w_s = w_0 = \sqrt{\dfrac{3\alpha}{4\beta} a_{0}}.
\end{equation}
Upon expansion, we realize that  three of the roots must be discarded since two of them do not agree with the classical result, and one yields $-w_0$. The correct root  proves to be
\begin{equation} \label{ws_root}
 w_s =  \dfrac{\sqrt{30}}{40 \sqrt{\beta K_1}} \sqrt{\alpha K_2 - \sqrt{K_3 \alpha^2+K_4 \alpha + 49} + 7},
\end{equation}
and its expansion can be found
\begin{equation} \label{ws}
\boxed{
    w_s = w_0 \left( 1 + \dfrac{1}{7} a_0 \overline{B} \alpha - \dfrac{1}{98} a_0^2 \overline{B}^2 \alpha^2 \right),}
\end{equation}
where we have defined
\begin{equation}\label{bbar}
    \overline{B} = 2B_1 + 90 B_2 -12 B_3 + 9B_4.
\end{equation}
By substituting equation \eqref{ws} into the formula for the velocity \eqref{V_s_w_s}, we finally obtain
\begin{equation} \label{V}
\boxed{
    V_s = 1 + \dfrac{1}{2} a_0 \alpha + \left(\dfrac{4}{35} a_0^2 B_1 + \dfrac{1}{35} a_0^2 \overline{B} - \dfrac{38}{35} a_0^2 B_4 + \dfrac{5}{7} a_0^2 B_3 - \dfrac{27}{7} a_0^2 B_2\right) \alpha^2. }
\end{equation}
For future reference, we denote the coefficient of the second-order term in the previous expression by $\overline{V}$:
\begin{equation*} 
    \overline{V} = \dfrac{4}{35} a_0^2 B_1 + \dfrac{1}{35} a_0^2 \overline{B} - \dfrac{38}{35} a_0^2 B_4 + \dfrac{5}{7} a_0^2 B_3 - \dfrac{27}{7} a_0^2 B_2.
\end{equation*}

%\textbf{Remark 3}. In the cases where we might expect that the changes in the width of the soliton, as it propagates, can be reliably predicted by the classical theory, that is   $w_s = \sqrt{\dfrac{3}{4} \dfrac{ a_0\alpha}{ \beta} }$ [Whitham], we can substitute this into the expression for the velocity, which produces
 %\begin{align*}
  %   V_s = 1 + \dfrac{1}{2}a_0 \alpha + \left( \dfrac{4}{35} B_1 a_0^2 - \dfrac{152}{105} B_4 a_0^2 + \dfrac{94}{105} B_3 a_0^2 - \dfrac{27}{7} B_2 a_0^2 \right) \alpha^2 + \left( \dfrac{19 \sqrt{3}}{105} B_4 a_0^{\frac{5}{2}} - \dfrac{19 \sqrt{3}}{210} B_3 a_0^{\frac{5}{2}}\right) \alpha^{\frac{5}{2}}\beta^{\frac{-1}{2}}
 %\end{align*}

\subsection{Energy method: eKdV velocity-amplitude relation}\label{subsubsec:3.2}
Now to find an explicit expression for the actual (higher order) solitonic height, one can obtain an additional and distinct velocity-amplitude relation by deriving an energy estimate for the eKdV equation (\ref{eKdV}). We refer to \cite{patkawahara} and \cite{hamidgk} for the application of the method to the Kawahara equation and to the conservative-eKdV equation, respectively. To start, we assume a general travelling wave solution in the form $u=u(x-V_{s}t)$ and substitute it into equation (\ref{eKdV}). This implies
\begin{equation}
    -V_{s}u'+ u'+ \dfrac{3}{4}(u^2)' \alpha + \dfrac{1}{6 }u^{(3)} \beta + \frac{1}{3} B_1 (u^3)' \alpha^2 + B_2 u^{(5)} \beta^2 + \left( \frac{1}{2} B_3(u'^2)' +B_4 uu^{(3)} \right) \alpha \beta =0,
\end{equation}
and by integrating the above equation we get 
\begin{equation}\label{prior_energy}
    -V_{s}u+ u + \dfrac{3}{4}u^2 \alpha + \dfrac{1}{6} u'' \beta + \frac{1}{3}B_1 u^3 \alpha^2 + B_2 u^{(4)} \beta^2 + \left( \frac{1}{2} (B_3-B_4) u'^2 + B_4 u u'' \right) \alpha \beta =E_{1},
\end{equation}
where $E_{1}$ is a constant of integration that can be determined by the boundary condition 
\begin{equation}
    \lim_{\theta\to{-\infty}}u= 0 \quad \implies \quad E_{1}= 0.
\end{equation}
Now, we multiply equation (\ref{prior_energy}) by $u'$ and integrate, giving the energy balance
\begin{multline}\label{energy_integral_1}
    -\frac{V_{s}}{2}u^2 + \dfrac{1}{2} u^2 + \dfrac{1}{4 }u^3 \alpha +\frac{1}{12} u'^2 \beta + \frac{ 1}{12} B_1 u^4 \alpha^2 + B_2 \left(u'u^{(3)} -\frac{1}{2}u''^2\right) \beta^2 \\
    + \left( \frac{ B_3-B_4}{2} \int u'^2 u'  + B_{4} \int u u' u'' \right) \alpha \beta = E_{2},
\end{multline}
where again the constant $E_{2}$ can be determined similarly by the boundary condition 
\begin{equation}
     \lim_{\theta\to{-\infty}}u= 0 \quad \implies \quad E_{2}= 0.
\end{equation}
On using integration by parts, one can put the integrals in (\ref{energy_integral_1}) in terms of the other:
\begin{multline}\label{energy_integral_2}
    -\frac{V_{s}}{2}u^2 + \dfrac{1}{2} u^2 + \dfrac{1}{4 }u^3 \alpha +\frac{1}{12} u'^2 \beta + \frac{ 1}{12} B_1 u^4 \alpha^2 + B_2 \left(u'u^{(3)} -\frac{1}{2}u''^2\right) \beta^2\\
    + \left( \frac{ B_3-B_4}{2} u'^2 u  + (2B_{4}-B_3) \int u u' u'' \right) \alpha \beta = 0.
\end{multline}
Equation (\ref{energy_integral_2})  cannot be additionally integrated to obtain an exact differential equation. This is expected, and it is connected to the fact that the eKdV equation does not conserve the kinetic energy quantity $u^2/2$. However, one can derive a local energy equation accurate to second-order \cite{salehekdv}. This is achieved by utilizing the classical KdV equation \textit{asymptotically}, that is equation (\ref{eKdV}) up to $\mathcal{O}(\alpha, \beta)$ only 
\begin{align} \label{asymp}
(2B_4-B_3) \alpha \beta \int u u' u'' & \sim \frac{(2B_{4}-B_{3})}{6}\int \left( \dfrac{2V_{s}}{3}u' \beta - \dfrac{2}{3} u' \beta - \dfrac{1}{9} u^{(3)} \beta^2 \right) u''  \\
& \sim \frac{2B_{4}-B_{3}}{3}\left( V_{s}u'^2 \beta - u'^2 \beta - \dfrac{1}{6}u''^2 \beta^2 \right),
\end{align}
which leads to the asymptotic energy equation 
\begin{multline}\label{energy_integral_approx}
    -\frac{V_{s}}{2}u^2 + \dfrac{1}{2} u^2 + \dfrac{1}{4 }u^3 \alpha +\left( \frac{1}{12} u'^2 + \dfrac{(2B_4-B_3)V_s}{3} u'^2 -\dfrac{2B_4-B_3}{3} u'^2\right) \beta \\
    + \frac{ 1}{12} B_1 u^4 \alpha^2 +  \left( B_2 u'u^{(3)} - \frac{B_2}{2}u''^2 - \dfrac{2B_4-B_3}{18} u''^2 \right) \beta^2
    + \left( \frac{ B_3-B_4}{2} u'^2 u  \right) \alpha \beta = 0
\end{multline}
We now substitute the ansatz of a solitary wave solution, travelling on zero background with the actual solitonic height $A_{s}$,   
\begin{equation}
    u= A_s \mathrm{ sech}^2(w_s \theta)
    \label{e:single_soliton}
\end{equation}
into (\ref{energy_integral_approx}) and evaluate at the maximum solitary wave amplitude $\theta=0$. This gives
\begin{multline}\label{energy_max}
    -\frac{V_{s}}{2}u_{\rm max}^2 + \dfrac{1}{2} u_{\rm max}^2 + \dfrac{1}{4 }u_{\rm max}^3 \alpha +\left( \frac{1}{12} u_{\rm max}'^2 + \dfrac{(2B_4-B_3)V_s}{3} u_{\rm max}'^2 -\dfrac{2B_4-B_3}{3} u_{\rm max}'^2\right) \beta + \\
    \frac{ B_1}{12}  u^4_{\rm max} \alpha^2 +  \left( B_2 u_{\rm max}'u_{\rm max}^{(3)} - \frac{B_2}{2}u_{\rm max}''^2 - \dfrac{2B_4-B_3}{18} u_{\rm max}''^2 \right) \beta^2
    + \left( \frac{ B_3-B_4}{2} u_{\rm max}'^2 u_{\rm max}  \right) \alpha \beta = 0,
\end{multline}
where 
\begin{equation}
    u_{\text{max}}= A_s,\quad u''_{\text{max}}= 2A_{s}w^{2}_{s},\quad u'_{\text{max}}=u^{(3)}_{\text{max}}=0.
\end{equation}
We now acquire a quadratic equation in $A_s$, which has the solutions
\begin{equation} \label{Hs_root}
    A_s = \dfrac{-9 \pm \sqrt{J_1 \beta^2 + J_2 + 81}}{6B_1 \alpha},
\end{equation}
where
\begin{equation}
    J_1 = 96 B_1 w_s^4 K_1, \quad J_2 = 216 B_1 (V_s-1).
\end{equation}
By stipulating that $A_s$ at zeroth order is identified with $a_0$, as expected, and because we consider a positive-polarity (bright or elevation) solitary wave solution in the application in the next section, we restrict our attention to the positive root in equation \eqref{Hs_root}. By employing the formulas for $w_s$ and $V_s$ found in the previous section, we can express the orders of $A_s$ in terms of the fixed amplitude $a_0$, yielding 
%However, we note that the asymptotic assumption in the relations given in equation \eqref{asymp} limits us to first order terms in both the velocity and width: $V_s = 1 + \dfrac{a}{2} \alpha$, and $w_s = w_0$. Hence, upon expansion we obtain
%\begin{equation} \label{Hs}
 %   H_s = a_0 + \dfrac{1}{2} a_0^2 K_1 \alpha - \dfrac{1}{3} a_0^3 B_1 K_1 \alpha^2
%\end{equation}
%Alternatively, we can insert the full expressions of $V_s$ and $w_s$ to obtain upon expansion
\begin{equation} \label{Hs-blue}
\boxed{
    A_s = a_0 + \left( 2\overline{V} + \dfrac{1}{2} a_0^2 K_1 - \dfrac{1}{3} a_0^2  B_1 \right) \alpha + \left( \dfrac{2}{7} a_0^3 \overline{B} K_1  - \dfrac{4}{3}  a_0  B_1 \overline{V} - \dfrac{1}{3} a_0^3 B_1 K_1 \right) \alpha^2.}
\end{equation}

\section{Application to a dispersive hydrodynamic problem}\label{sec:4}

\begin{figure}[!htp]
    \centering
\includegraphics[width=0.49\textwidth]{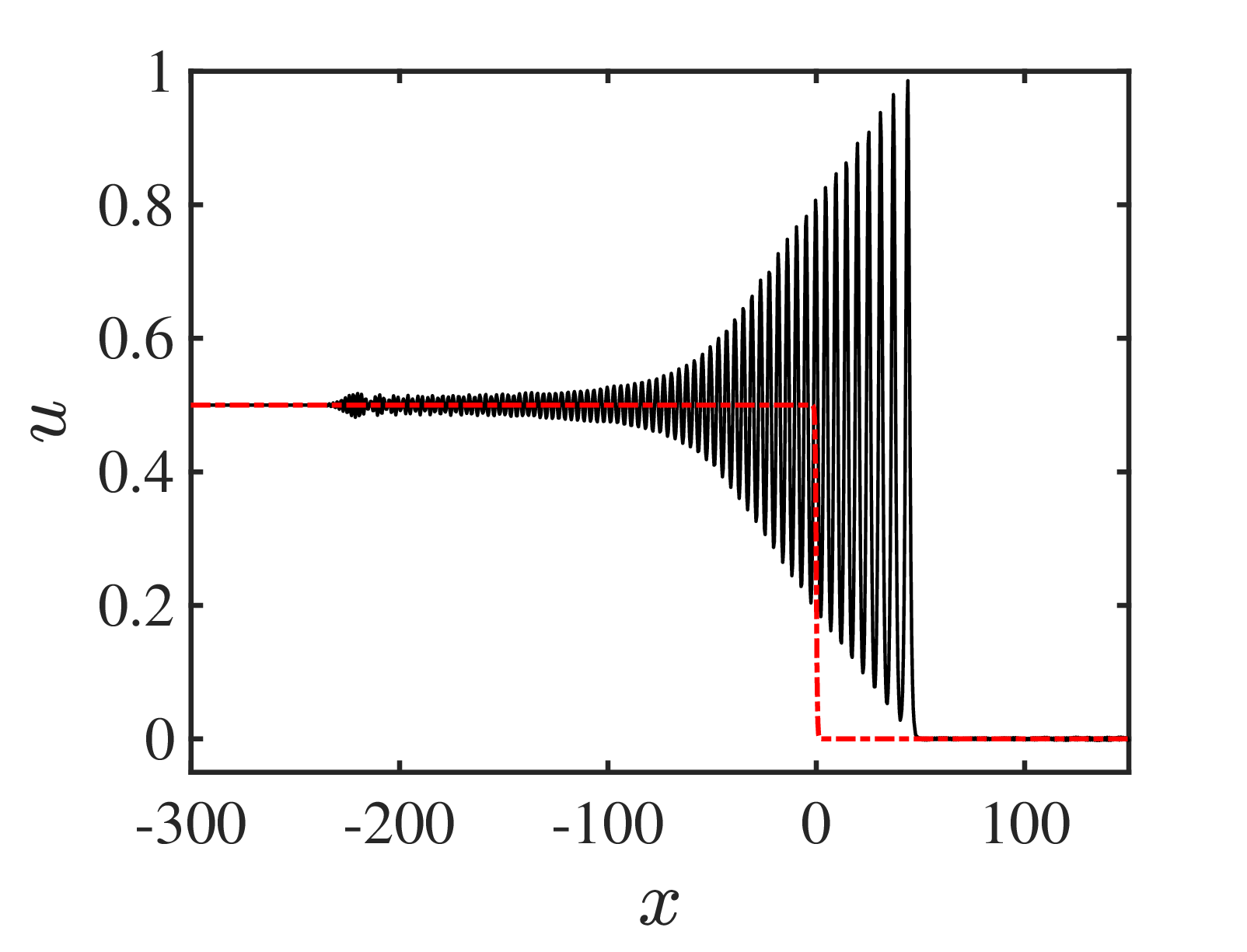} \includegraphics[width=0.49\textwidth]{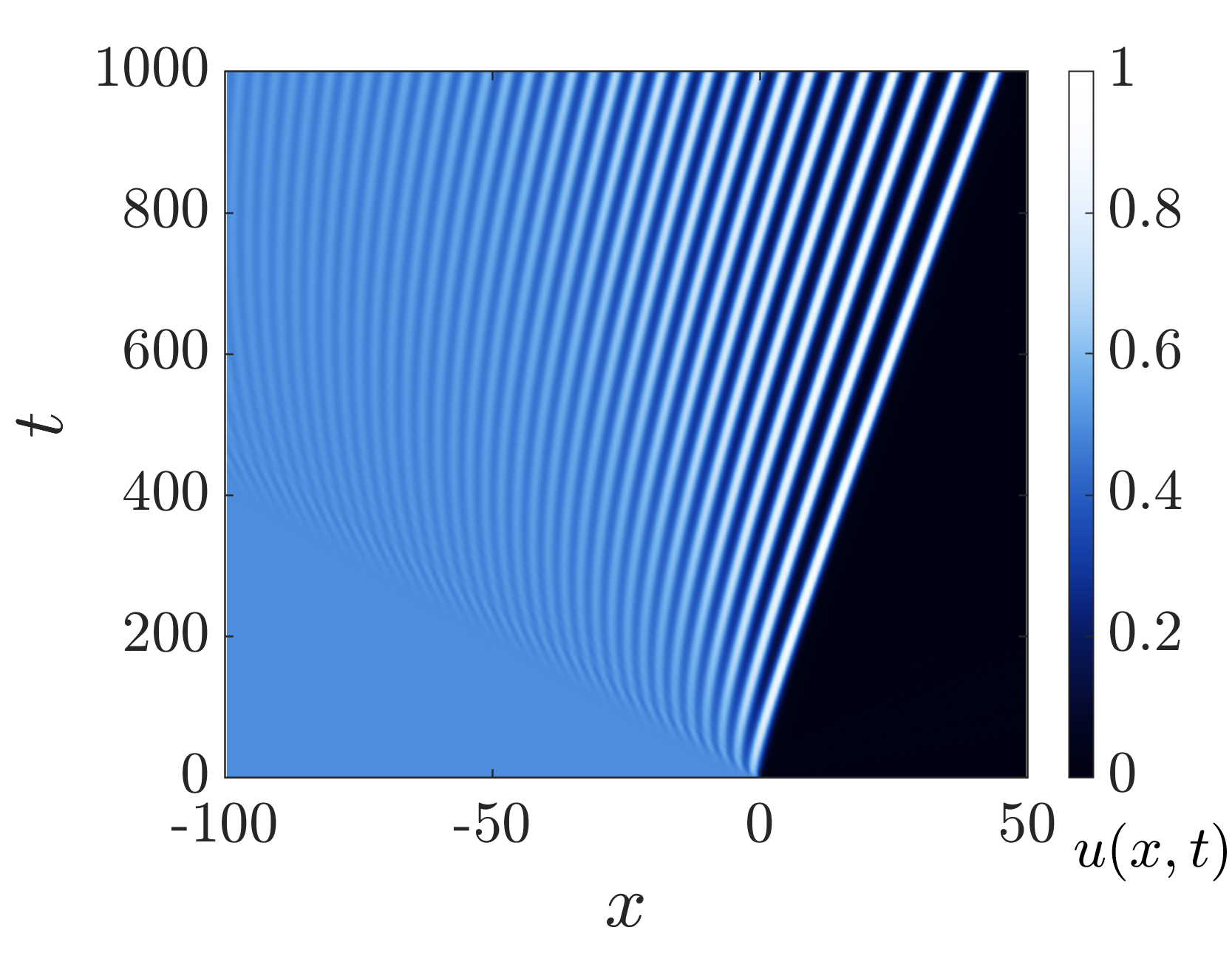}
    \caption{Dispersive shock regime governed by the eKdV equation \eqref{eKdV_transport}, together with a space–time contour plot of a portion of the solution $u(x,t)$ over the interval $x\in[-100,50]$. The red (dashed) line indicates the initial jump discontinuity \eqref{e:ic_eq}. The coefficients $B_{1}$, $B_{2}$, $B_{3}$, and $B_{4}$ are chosen as the shallow water wave coefficients \eqref{coefs}. Here, $\alpha = \beta = 0.1$, $\Delta = 0.5$, and $t = 1000$. (Color version online).}
     \label{fig:ekdv_dsw_plots}
\end{figure}

In this section, we consider a problem in dispersive hydrodynamics that necessitates the availability of a solitary wave solution of the full eKdV equation \eqref{eKdV}. We shall study the Riemann problem for the eKdV equation \eqref{eKdV}, which gives rise to a DSW, namely, a non-stationary modulated nonlinear periodic wavetrain connecting two distinct initial flow states. Figure \ref{fig:ekdv_dsw_plots} provides an illustration of an eKdV DSW.

Generally speaking, solutions describing dispersive shocks are obtained using Whitham modulation theory \cite{elreview,whitham,kamchatnovbook}, which is based on averaged conservation laws or, equivalently, averaged Lagrangian. However, the determination of exact modulation equations (a system of PDEs governing the slowly varying wave parameters of the underlying wavetrain) requires that the modulation system to be expressed in Riemann invariant form. This formulation is possible only when the nonlinear dispersive wave model under consideration is integrable \cite{analysis}.
As previously stressed, the full eKdV equation \eqref{eKdV} is a non-integrable model. Consequently, alternative approximate methods are required to determine the associated DSW solutions. 

One particularly powerful approach in dispersive hydrodynamics—and a central focus of the present work—is the DSW equal amplitude approximation developed by Smyth and Marchant in \cite{equalamp}. This method enables the determination of certain macroscopic properties of DSWs, notably the height and velocity of the leading solitary wave edge, without requiring full knowledge of the associated modulation equations. Importantly, the method is applicable regardless of whether the governing equation is integrable. A wide range of applications to nonlinear dispersive wave problems can be found in \cite{salehekdv,salehnem2,salehnem1}. It is also worth mentioning another related approach, the DSW fitting method developed by El \cite{fitting}. This method can provide predictions for a broader range of macroscopic and microscopic DSW properties, again without requiring explicit knowledge of the full modulation system. However, we do not consider the fitting method further here, as it is not directly relevant to the present work. In particular, unlike the equal amplitude approximation, the fitting method does not require the explicit existence of a solitary wave solution in its formulation. Interested readers can refer to \cite{elreview} for further details and applications of the DSW fitting method to both integrable and non-integrable dispersive systems.

The DSW equal amplitude approximation for dispersive shocks is typically applied to nonlinear dispersive wave equations in a moving reference frame. Specifically, we apply a Galilean transformation, $\xi=x-t$ and $\tau=t$, to the eKdV equation (\ref{eKdV}), which eliminates the transport term $u_{x}$. The eKdV equation then takes the form (after replacing $\xi$ and $\tau$ by $x$ and $t$, respectively):
\begin{equation}\label{eKdV_transport}
    u_{t} +\dfrac{3}{2} uu_{x} \alpha + \dfrac{1}{6}u_{3x} \beta +B_{1}u^2u_{x} \alpha^2 +B_{2}u_{5x} \beta^2 + \left( B_{3}u_{x}u_{xx}+B_{4}uu_{3x} \right) \alpha \beta =0.
\end{equation}
In this case, the eKdV equation (\ref{eKdV_transport}) has the mass conservation equation 
\begin{equation}
 \frac{\partial}{\partial t} u + \frac{\partial}{\partial x} \left[  \frac{3}{4}\alpha u^{2} + \frac{1}{6}\beta u_{xx} + \frac{1}{3}\alpha^2 B_{1} u^{3} + B_{2}\beta^{2} u_{4x} + \alpha\beta B_{4}uu_{xx} + \frac{1}{2}\alpha\beta\left(B_3-B_4\right) u^2_{x} \right] = 0 .
 \label{e:ekdvmass}
\end{equation}
On multiplying the eKdV equation (\ref{eKdV_transport}) by $u$ and integrating by parts gives
\begin{eqnarray} 
 & & \frac{\partial}{\partial t} \left[\frac{1}{2}u^{2}\right] + \frac{\partial}{\partial x} \left[ \frac{1}{2}\alpha u^{3} + \frac{1}{6}\beta uu_{xx} -\frac{1}{12}\beta u^{2}_{x} +\frac{1}{4}\alpha^2 B_{1}u^{4} + \frac{1}{2}\alpha\beta B_{3}u^{2}u_{xx} + \beta^{2}B_{2}uu_{4x} \right. \nonumber \\
 & & \left.  \mbox{} - \beta^{2}B_{2}u_{x}u_{3x} + \frac{1}{2}\beta^2B_{2}u^{2}_{xx} \right] + \alpha\beta \left( B_{4} - \frac{1}{2} B_{3} \right) u^{2}u_{3x} = 0.
  \label{e:energycons}
\end{eqnarray}
To set the final term in conservation form we use the fact that $\alpha$ and $\beta$ are of small magnitude parameters. The standard KdV equation (\ref{KdV3-intro}) gives 
\begin{equation}
 \frac{\partial}{\partial t} u^{3} +\frac{9}{8}\alpha \frac{\partial}{\partial x}u^{4} = -\frac{1}{2}\beta u^{2}u_{3x},
 \end{equation}
so that the above unconserved form (\ref{e:energycons}) becomes the eKdV energy conservation equation
\begin{eqnarray}
 & & \frac{\partial}{\partial t} \left[\frac{1}{2}u^{2} - \alpha(2B_4-B_{3})u^{3} \right] + \frac{\partial}{\partial x} \left[ \frac{1}{2}u^{2} +\frac{1}{2}\alpha u^{3} + \frac{1}{6}\beta uu_{xx} -\frac{1}{12}\beta u^{2}_{x} +\frac{1}{4}\alpha^2 B_{1}u^{4} \right. + \nonumber \\
 & & \left.  \mbox{} \frac{1}{2}\alpha\beta B_{3}u^{2}u_{xx} + \beta^{2}B_{2}uu_{4x} - \beta^{2}B_{2}u_{x}u_{3x} + \frac{1}{2}\beta^2B_{2}u^{2}_{xx}  - \frac{9}{8}\alpha^2\left(2B_{4}-B_{3}\right)u^{4} \right] =0, 
  \label{e:energyconsf}
\end{eqnarray}
which is valid asymptotically.  We note that if $B_{3} = 2B_{4}$, then this energy conservation law is exact. %For further applications of derived single solitary wave solutions in the context of dispersive shock problems, particularly in the case of energy conservation, please refer to the following study [Ref].

Let us now consider the eKdV Riemann problem, namely, the nonlinear dispersive wave equation (\ref{eKdV_transport}) subject to the initial jump 
\begin{equation}
    u(x,0)= \left\{ \begin{array}{cc}
         \Delta,\quad x<0\\
          0,\quad x>0,
    \end{array}\right.
    \label{e:ic_eq}
\end{equation}
which generates a dispersive shock. The parameter $\Delta$ represents the magnitude of the initial discontinuity. To generate a shock, we need to impose $\Delta>0$. To implement the DSW equal amplitude approximation, we assume that the DSW primarily consists of a train of $\mathcal{N}$ solitary waves with nearly equal amplitudes, and approximately uniformly distributed throughout the DSW. Then we start by integrating the above conservation laws \eqref{e:ekdvmass} and \eqref{e:energyconsf} over the DSW domain $-\infty<x<\infty$. This leads to the averaged equations upon using the boundary conditions at $x = \pm  \infty$
\begin{equation}
   \mathcal{N}\frac{d}{dt}\left(\overline{u}\right)= \frac{3}{4}\alpha\Delta^{2}+\frac{1}{3}\alpha^2B_{1}\Delta^{3},
   \label{e:aveq1}
\end{equation}
\begin{equation}
   \mathcal{N}\frac{d}{dt}\left(\frac{1}{2}\overline{u^{2}} - \alpha(2B_4-B_{3})\overline{u^{3}}\right)=  \frac{1}{2}\alpha\Delta^{3}+\left(\frac{1}{4}B_{1}-\frac{9}{8}\left(2B_{4}-B_{3}\right)\right)\Delta^{4}\alpha^{2}-\alpha\left(2B_{4}-B_{3}\right)\Delta^{3},
   \label{e:aveq2}
\end{equation}
Using the earlier derived variational single solitary wave solutions  \eqref{ws}, \eqref{V} and \eqref{Hs-blue}, we can calculate the averaged mass and energy densities (up to second-order in the parameter $\alpha$)
\begin{align}
     \overline{u} & =2\dfrac{a_{0}}{w_{0}}-\dfrac{22}{105}\dfrac{a^{2}_{0}}{w_{0}}\left(B_{1}+\dfrac{810}{11}B_{2}-\dfrac{150}{11}B_{3}+\dfrac{228}{11}B_{4}+\dfrac{9}{11}\overline{B} -\dfrac{105}{22}K_{1} \right)\alpha \nonumber \\
 & \mbox{} ~~~ - \dfrac{32}{105}\dfrac{a^{3}_{0}}{w_{0}}\left(B^{2}_{1}  + \left[\dfrac{35}{16}K_{1}+\dfrac{17}{112}\overline{B} -\dfrac{19}{2}B_{4}+\dfrac{25}{4}B_{3}-\dfrac{135}{4}B_{2}\right]B_{1} -\dfrac{405}{56}\left[ B_{2} -\dfrac{5}{27}B_{3} \right. \right. \nonumber \\
  & \mbox{} ~~~ \left.\left.  +\dfrac{38}{135}B_{4} +\dfrac{1}{540}\overline{B}+\dfrac{36}{7}K_{1}\right]\overline{B}\right)\alpha^{2}+\mathcal{O}\left(\alpha^{3}\right),
\end{align}
\begin{align}
    \overline{u^{2}} & = \dfrac{4}{3}\dfrac{a^{2}_{0}}{w_{0}}-\dfrac{88}{315}\dfrac{a^{3}_{0}}{w_{0}}\left(B_{1}+\dfrac{810}{11}B_{2}-\dfrac{150}{11}B_{3}+\dfrac{228}{11}B_{4}+\dfrac{3}{22}\overline{B}-\dfrac{105}{22}K_{1}\right)\alpha \nonumber \\
   &  \mbox{} ~~~ -\dfrac{12596}{33075 }\dfrac{a^{4}_{0}}{w_{0}}\left(B^{2}_{1} -\dfrac{656100}{3239} B^{2}_{2} - \dfrac{22500}{3239}B^{2}_{3} -\dfrac{51984}{3239}B^{2}_{4}+\dfrac{63}{6478}\overline{B}^{2} -\dfrac{11025}{12956}K^{2}_{1} \right. \nonumber \\ 
   &  \mbox{} ~~~ \left. -\dfrac{5355}{3239}\overline{B}K_{1} +\bigg[ 8505 K_{1}- 131220 B_{2}+ 24300 B_{3}-36936 B_{4}+642\overline{B}\bigg]\dfrac{B_{1}}{3239} \right. \nonumber \\
   &  \mbox{} ~~~ \left. + \bigg[ 243000 B_{3}- 369360 B_{4} -14580 \overline{B} + 85050 K_{1} \bigg]\dfrac{B_{2}}{3239} + \bigg[ 23940K_{1} -4104\overline{B} \bigg]\dfrac{B_{4}}{3239} \right. \nonumber \\
   &  \mbox{} ~~~ \left. + \bigg[ 68400 B_{4} + 2700 \overline{B} - 15750 K_{1} \bigg] \dfrac{B_{3}}{3239} \right)\alpha^{2} +\mathcal{O}\left(\alpha^{3}\right)
\end{align}
and 
\begin{align}
    \overline{u^3} & = \dfrac{16}{15}\dfrac{a^{3}_{0}}{w_{0}} - \dfrac{176}{525}\dfrac{a^{4}_{0}}{w_{0}}\left( B_{1} +\dfrac{810}{11}B_{2} -\dfrac{150}{11}B_{3} + \dfrac{228}{11}B_{4} -\dfrac{1}{11}\overline{B} -\dfrac{105}{22}K_{1} \right) \nonumber \\
   &  \mbox{} ~~~ -\dfrac{24944}{55125 }\dfrac{a^{5}_{0}}{w_{0}}\left(B^{2}_{1} -\dfrac{656100}{1559} B^{2}_{2} - \dfrac{22500}{1559}B^{2}_{3} -\dfrac{51984}{1559}B^{2}_{4}+\dfrac{33}{3118}\overline{B}^{2} -\dfrac{11025}{6236}K^{2}_{1} \right. \nonumber \\ 
   &  \mbox{} ~~~ \left. -\dfrac{5985}{3118}\overline{B}K_{1} +\bigg[ 4830 K_{1}- 74520 B_{2}+ 13800 B_{3}-20976 B_{4}+ 387\overline{B}\bigg]\dfrac{B_{1}}{1559} \right. \nonumber \\
   &  \mbox{} ~~~ \left. + \bigg[ 243000 B_{3}- 369360 B_{4} -2430 \overline{B} + 85050 K_{1} \bigg]\dfrac{B_{2}}{1559} + \bigg[ 23940K_{1} -684\overline{B} \bigg]\dfrac{B_{4}}{1559} \right. \nonumber \\
   &  \mbox{} ~~~ \left. + \bigg[ 68400 B_{4} + 450 \overline{B} - 15750 K_{1} \bigg] \dfrac{B_{3}}{1559} \right)\alpha^{2} +\mathcal{O}\left(\alpha^{3}\right)
\end{align}
with $w_{0}$ and $\overline{B}$ given by the expressions \eqref{w0} and \eqref{bbar}. Upon integrating the averaged equations (\ref{e:aveq1}) and (\ref{e:aveq2}) with respect to time and dividing one by the other, we obtain the relation  
\begin{equation}
  \frac{\overline{u}}{\dfrac{1}{2}\overline{u^{2}}-\alpha\left(2B_{4}-B_{3}\right)\overline{u^{3}}}=\dfrac{\dfrac{3}{4}\alpha\Delta^{2}+\dfrac{1}{3}\alpha^2B_{1}\Delta^{3}}{\dfrac{1}{2}\Delta^{2}+\dfrac{1}{2}\alpha\Delta^{3}+\left[\dfrac{1}{4}B_{1}-\dfrac{9}{8}\left(2B_{4}-B_{3}\right)\right]\Delta^{4}\alpha^{2}-\alpha(2B_{4}-B_{3})\Delta^{3}},
  \label{e:algebraiceq}
\end{equation}
by assuming that there are no solitary waves in the DSW at $t=0$. Equation (\ref{e:algebraiceq}) now becomes an algebraic equation in the unknown parameter $a_{0}$ which can be detremined numerically. Upon determining $a_{0}$, the height $A_{s}$ of the lead solitary wave edge of the DSW can be predicted using (\ref{Hs-blue}). Furthermore, the velocity $V_{s}$ of the lead solitary wave edge of the DSW can also be ascertained through the previously derived velocity-amplitude relation (\ref{V}), after accounting for the change in reference frame.

The variational single solitary wave solution derived for the full eKdV equation \eqref{eKdV} can likewise be applied to a variety of other problems in dispersive hydrodynamics. For example, it may be used to study non-classical DSW formations governed by the eKdV equation, including the so-called cross-over dispersive shock propagation. The authors have previously considered said application in the simpler case when $B_{3}=2B_{4}$ (the energy conservation case) in which the eKdV equation can be obtained by a considerably simpler Lagrangian \cite{hamidgk}.

\section{Numerical results, comparisons, and discussions}\label{sec:5}

This section is divided into two parts. First, we compare the variational solitary wave and DSW solutions derived above with direct numerical solutions of the eKdV equation \eqref{eKdV}, which are obtained using a pseudo-spectral method. In particular, the spatial domain is discretized using the Fast Fourier Transform (FFT), with the integrating factor method employed to treat the stiff terms, following the approach described in \cite{trefethen}. Temporal integration is then performed using the classical fourth-order Runge–Kutta (RK4) numerical scheme. In the second part of this section, we compare and discuss the variational solitary wave results obtained in this work with the algebraic solitary wave results reported in earlier studies by the authors of \cite{annabook,annapaper}. In particular, the constant value of the eKdV solitary wave velocity $V_s$ reported in \cite{annapaper}, which stands in contrast to the numerical solutions, is examined and its applicability is explained.

\begin{figure}[!htp]
\centering

\begin{subfigure}[b]{0.90\textwidth}
    \centering
    \begin{minipage}{0.07\textwidth}
        \caption{}
    \end{minipage}
    \hspace{0.02\textwidth}
    \begin{minipage}{0.80\textwidth}
        \includegraphics[width=\textwidth]{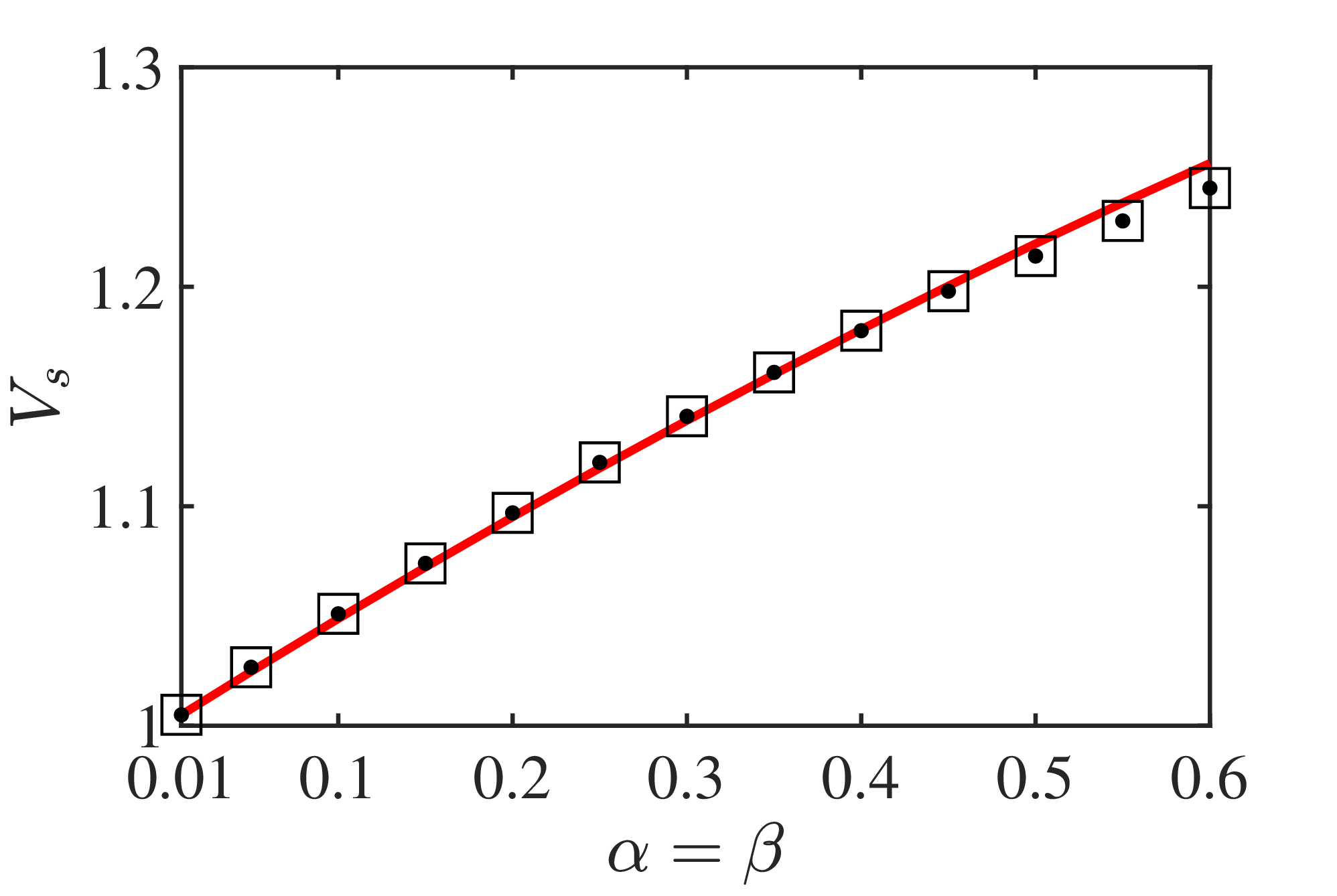}
    \end{minipage}
\end{subfigure}

\vspace{0.5cm}

\begin{subfigure}[b]{0.90\textwidth}
    \centering
    \begin{minipage}{0.07\textwidth}
        \caption{}
    \end{minipage}
    \hspace{0.02\textwidth}
    \begin{minipage}{0.80\textwidth}
        \includegraphics[width=\textwidth]{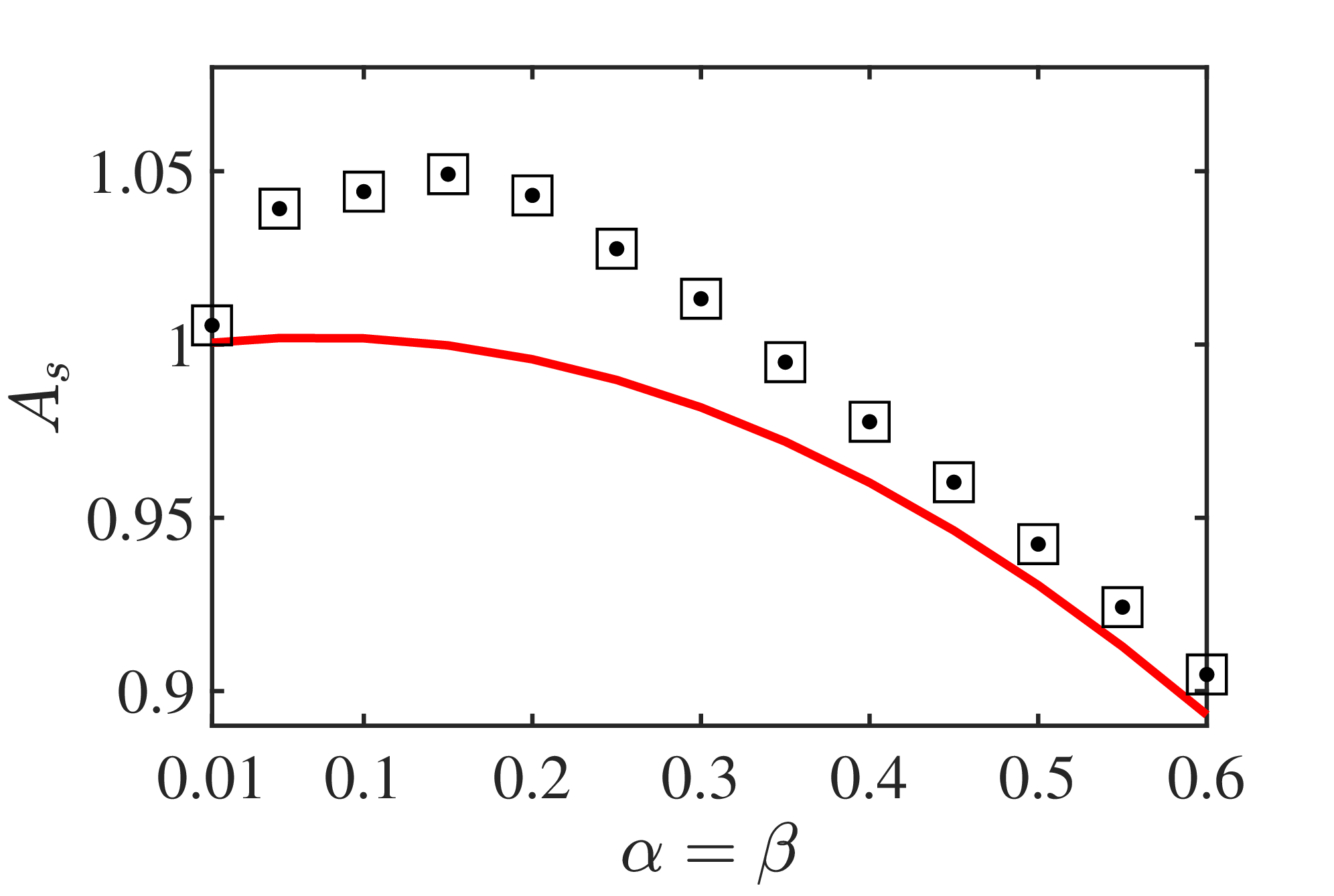}
    \end{minipage}
\end{subfigure}

\caption{Comparisons for solitonic wave parameters governed by the eKdV equation \eqref{eKdV}. Numerical solutions: black (dotted) box; theoretical solutions: red (solid) line. (a) comparison for eKdV solitary wave velocity, (b) comparison for eKdV solitary wave height. The coefficients $B_{1}$, $B_{2}$, $B_{3}$, and $B_{4}$ are taken to be the shallow water wave coefficients \eqref{coefs}. Here, $a_{0}=1$ and $t=40$. (Color version online).}
\label{fig:sol_comps}

\end{figure}

\begin{figure}[!htp]
\centering

\begin{subfigure}[b]{0.90\textwidth}
    \centering
    \begin{minipage}{0.07\textwidth}
        \caption{}
    \end{minipage}
    \hspace{0.02\textwidth}
    \begin{minipage}{0.80\textwidth}
        \includegraphics[width=\textwidth]{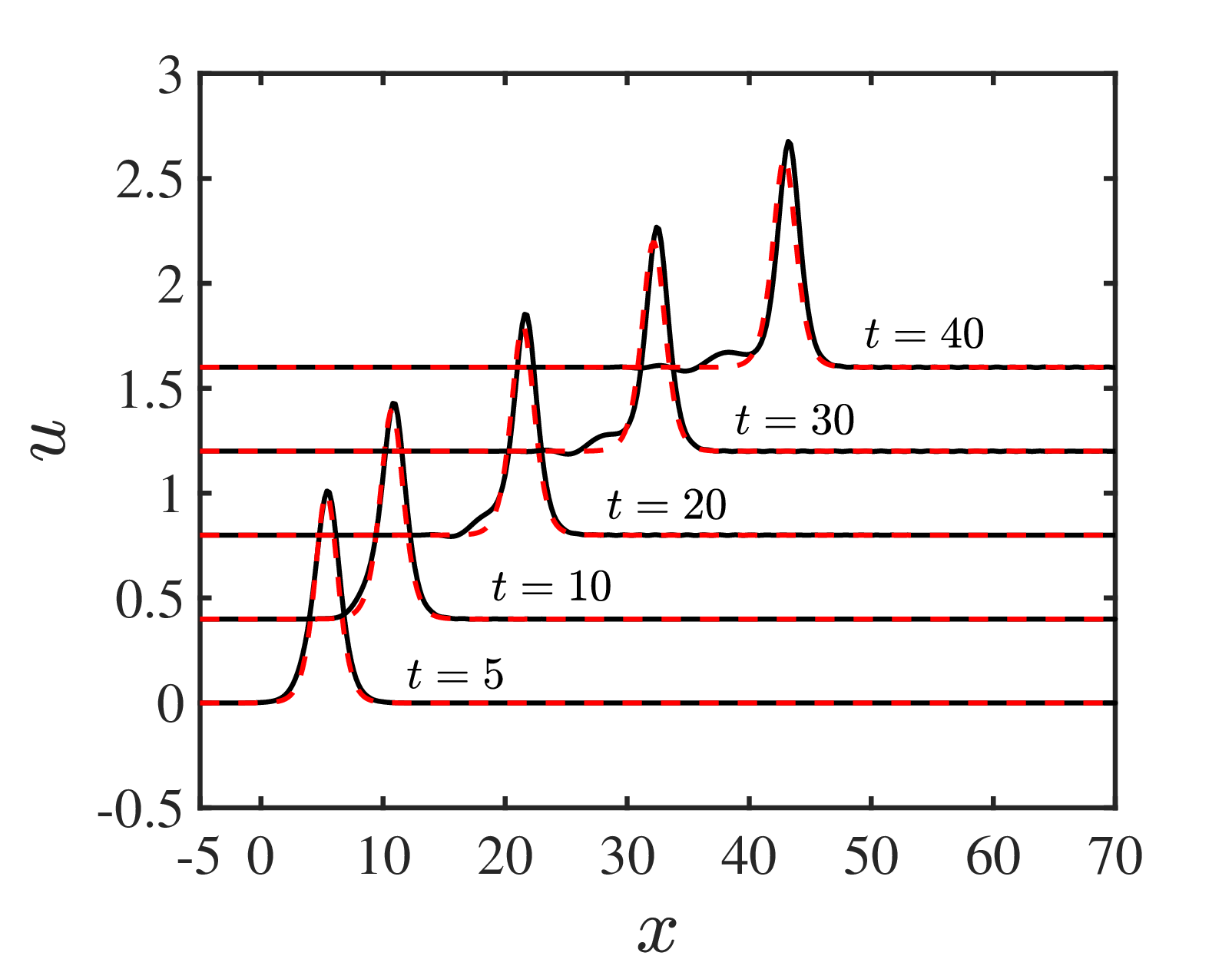}
    \end{minipage}
\end{subfigure}

\vspace{0.5cm}

\begin{subfigure}[b]{0.90\textwidth}
    \centering
    \begin{minipage}{0.07\textwidth}
        \caption{}
    \end{minipage}
    \hspace{0.02\textwidth}
    \begin{minipage}{0.80\textwidth}
        \includegraphics[width=\textwidth]{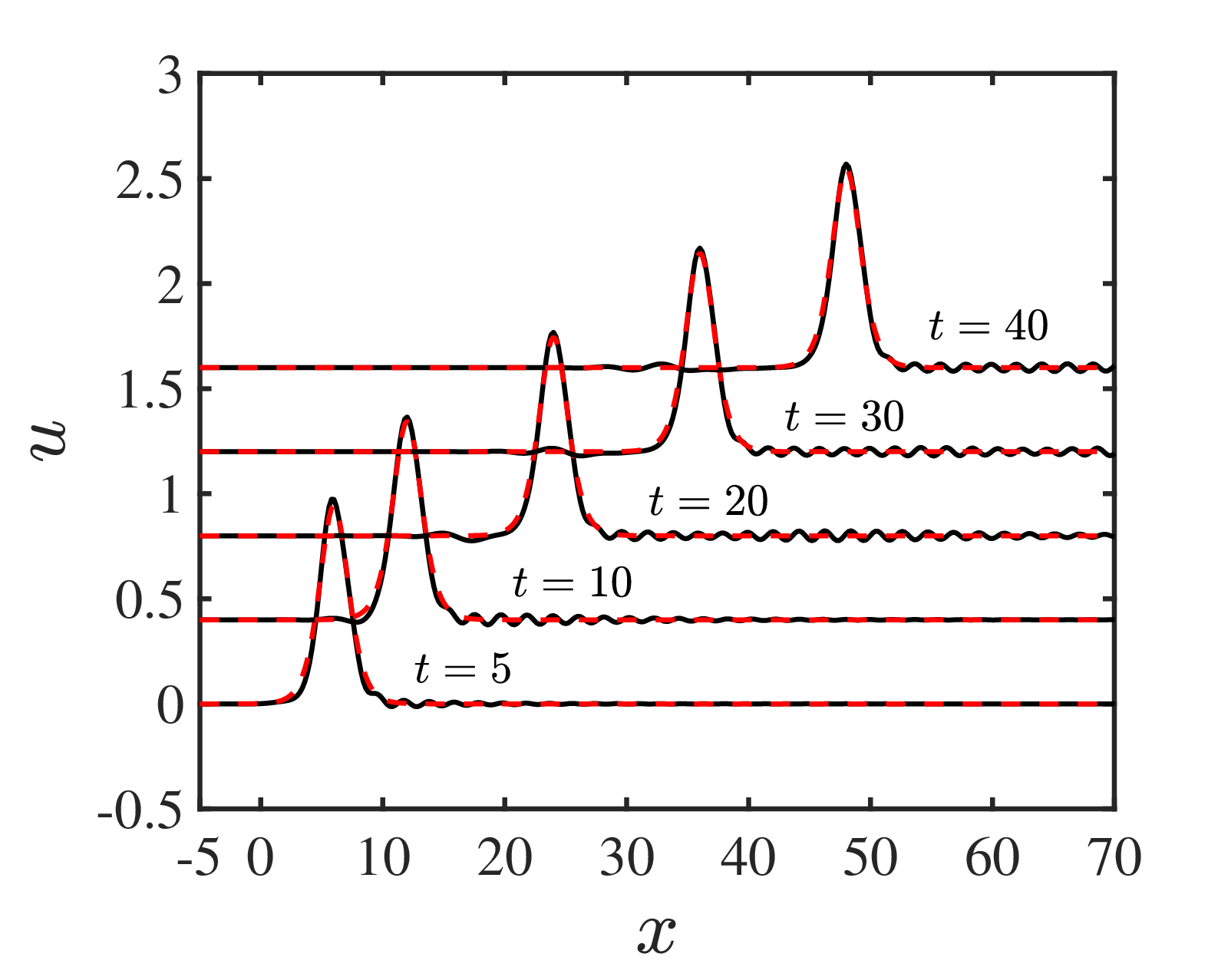}
    \end{minipage}
\end{subfigure}

\caption{Evolution of shallow water solitary waves governed by the eKdV equation \eqref{eKdV}. Numerical solutions: black (solid) line; theoretical solutions: red (dashed) line. In (a), the asymptotic parameters are set to $\alpha=\beta=0.15$. In (b), the asymptotic parameters are set to $\alpha=\beta=0.45$. Compared to (a), the solitary wave tail emits resonant radiation. The coefficients $B_{1}$, $B_{2}$, $B_{3}$, and $B_{4}$ are taken to be the shallow water wave coefficients \eqref{coefs}. Here, $a_{0}=1$. (Color version online).}
\label{fig:evolve_comps}

\end{figure}

\begin{figure}[!htp]
\centering

\begin{subfigure}[b]{0.90\textwidth}
    \centering
    \begin{minipage}{0.07\textwidth}
        \caption{}
    \end{minipage}
    \hspace{0.02\textwidth}
    \begin{minipage}{0.80\textwidth}
        \includegraphics[width=\textwidth]{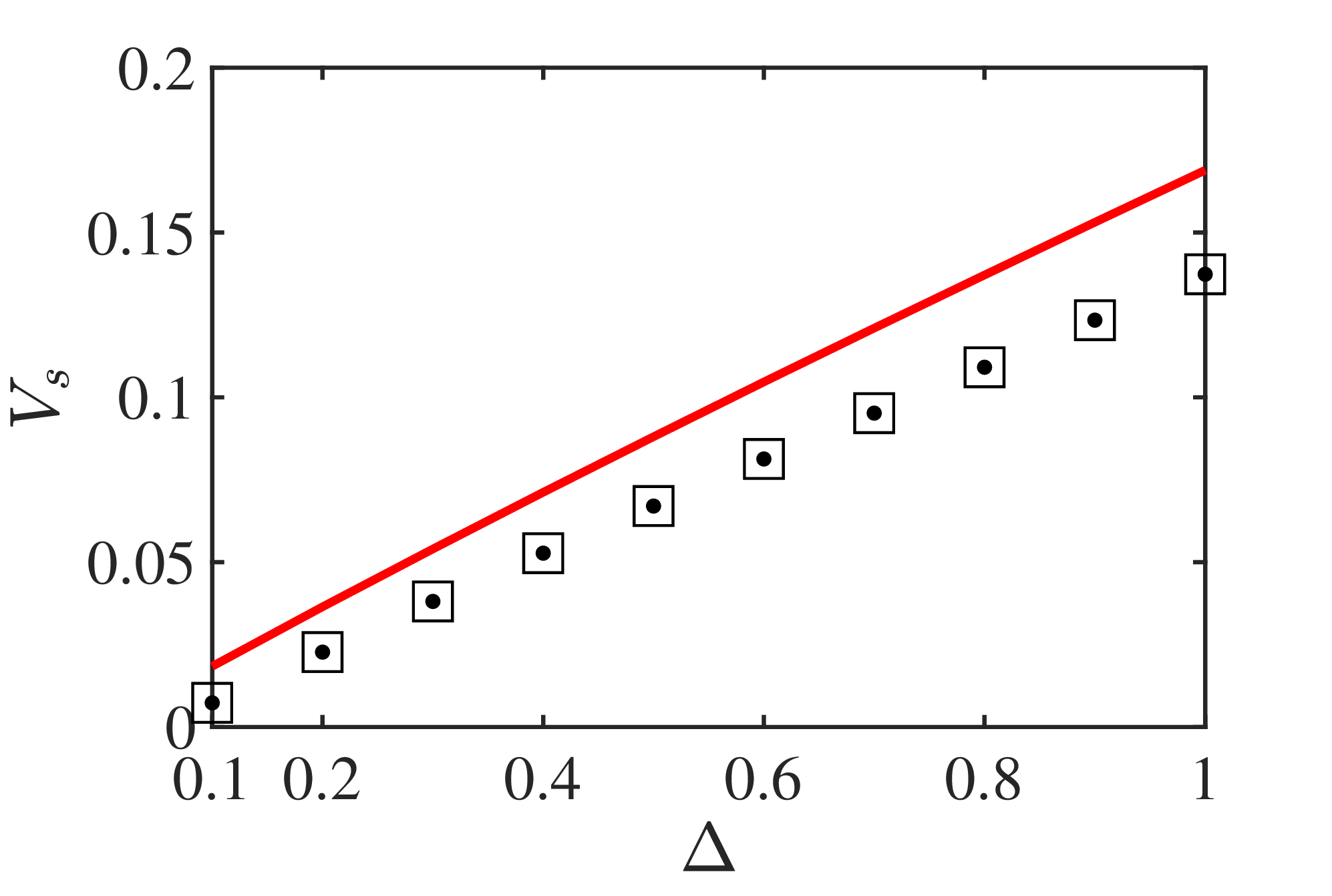}
    \end{minipage}
\end{subfigure}

\vspace{0.5cm}

\begin{subfigure}[b]{0.90\textwidth}
    \centering
    \begin{minipage}{0.07\textwidth}
        \caption{}
    \end{minipage}
    \hspace{0.02\textwidth}
    \begin{minipage}{0.80\textwidth}
        \includegraphics[width=\textwidth]{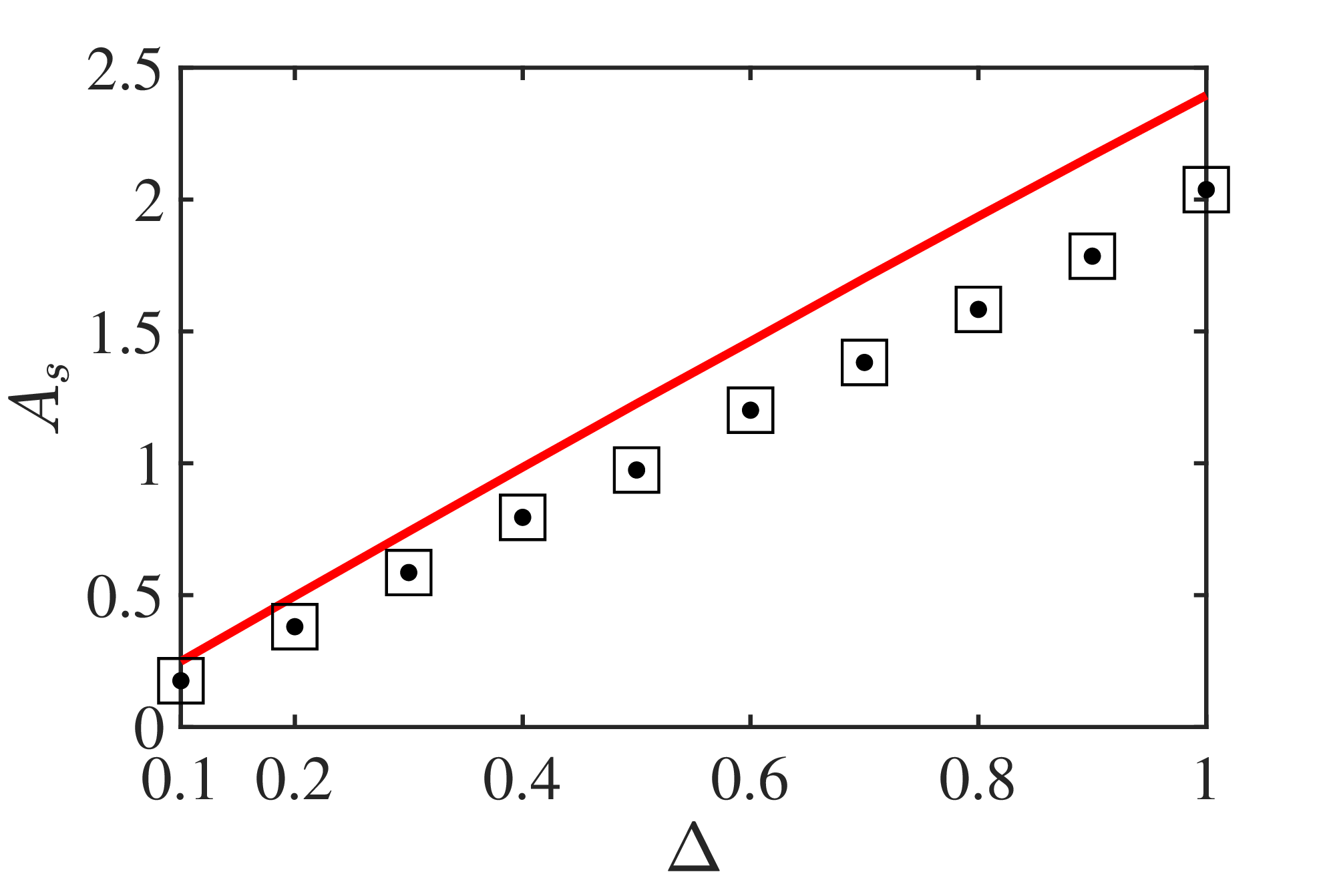}
    \end{minipage}
\end{subfigure}

\caption{Comparisons for dispersive shock parameters governed by the eKdV equation without the transport term $u_{x}$ \eqref{eKdV_transport}. Numerical solutions: black (dotted) box; theoretical solutions: red (solid) line. (a) comparison for the lead solitary wave edge velocity of eKdV DSW, (b) comparison for the lead solitary wave edge height of eKdV DSW. The coefficients $B_{1}$, $B_{2}$, $B_{3}$, and $B_{4}$ are chosen as the shallow water wave coefficients \eqref{coefs}. Here, $\alpha,\beta=0.15$ and $t=1000$. (Color version online).}
\label{fig:dsw_comps}

\end{figure}

\subsection{Comparisons for solitary waves and DSW solutions}

For simplicity, we set $\alpha=\beta$. Figures \ref{fig:sol_comps}(a) and \ref{fig:sol_comps}(b) present comparisons between the theoretical predictions and numerical results for the solitary wave velocity $V_{s}$ and height $A_{s}$ of the eKdV equation \eqref{eKdV} as functions of the parameters $\alpha$ and $\beta$, with $a_{0}=1$, over a wide parameter range satisfying the condition $\alpha,\beta<1$. The higher order coefficients $B_{1}$, $B_{2}$, $B_{3}$, and $B_{4}$ are taken to be the shallow water values given in \eqref{coefs}. Overall, the agreement between the theoretical predictions and the numerical computations is excellent. 

It is worth noting that, in the numerical profile of the solitary wave height, a maximum point is observed beyond which the height changes its monotonicity. This behavior is associated with the emergence of resonant radiation propagating ahead of the solitary wave as the parameters $\alpha=\beta$ increase, a phenomenon that is also evident in Figures \ref{fig:evolve_comps}(a) and \ref{fig:evolve_comps}(b). The resonant radiation extracts mass from the underlying solitary wave, thereby contributing to the subsequent decrease in the observed wave height. Despite this effect, the decreasing feature in the height is still captured well by the theoretical predictions. This is noteworthy because the variational solutions assume a pure (single) $\sech^{2}$ solitary wave profile and does not explicitly account for the oscillatory radiating tails.  As emphasized earlier, our primary interest lies in the pure $\sech^{2}$ solitary wave structure governed by the eKdV equation \eqref{eKdV}, whose analytical tractability plays an important role in the study of both classical and non-classical DSWs. A detailed analysis of the oscillatory tails generated by resonant radiation requires the use of exponential asymptotic techniques, as discussed for example in \cite{resekdv,exp1,exp2,exp3,exp4}, and lies beyond the scope of the present study. 

For further illustration, Figures \ref{fig:evolve_comps}(a) and \ref{fig:evolve_comps}(b)  display snapshots of the numerical evolution of shallow water eKdV solitary waves over the time interval $t=5$ to $t=40$, for $\alpha=\beta=0.15$ and $\alpha=\beta=0.45$, respectively. These simulations show that the magnitude of the resonant radiating tails increases as the parameters $\alpha=\beta$ become larger. Nevertheless, once the radiative component is disregarded for the reasons discussed above, the agreement between the theoretical predictions and the numerical wave profiles remains excellent.

Figures \ref{fig:dsw_comps}(a)  and \ref{fig:dsw_comps}(b)  present comparisons between the theoretical predictions and numerical results for the leading solitary wave edge velocity $V_{s}$ and height $A_{s}$ of the eKdV DSWs, as derived in Section \ref{sec:4}, versus a range of initial jump magnitudes $\Delta$. In these computations, the higher order coefficients $B_{1}$, $B_{2}$, $B_{3}$, and $B_{4}$ are taken to be the shallow water values given in \eqref{coefs}, with $\alpha=\beta=0.15$.
The initial jump discontinuity is implemented numerically using a smooth tanh-type profile in order to avoid numerical instabilities associated with gradient catastrophe. Specifically, the initial condition is given by
\begin{equation}
    u(x,0)=\dfrac{\Delta}{2}\left[\tanh\left(\dfrac{x+D}{W}\right)-\tanh\left(\dfrac{x}{W}\right)\right],
\end{equation}
where $W$ denotes the width of the initial well and $D$ represents the spatial location at which the initial jump transitions to the upstream state ahead of the shock (which is zero in the present case; see equation \eqref{e:ic_eq}). From both comparison plots, it can be clearly seen that the theoretical predictions are in excellent agreement with the numerical solutions for the entire range of jump magnitudes considered.

\subsection{Comparisons with previous works}
One observes from the variational solitary wave parameter solutions \eqref{ws}, \eqref{V}, and \eqref{Hs-blue} that they possess the important property of \emph{asymptotic  reduction}. By this we mean that, when the higher order effects (second-order approximation) in the eKdV equation are neglected, the solutions \eqref{ws}, \eqref{V}, and \eqref{Hs-blue} reduce directly to the corresponding (first-order) solutions associated with the classical KdV equation, namely
\begin{equation}\label{classical_1}
w_s = \sqrt{\frac{3 \alpha}{4\beta}a_0},\quad V_s = 1+ \frac{a_0}{2}\alpha,\quad A_{s}=a_{0}.  
\end{equation}
The availability of higher order solitonic solutions that retain this asymptotic reduction property is crucial for a broad range of applications in dispersive hydrodynamics; as illustrated in the present paper, in the authors’ previous work \cite{hamidgk}, and in \cite{tim_soliton}.

In \cite{annabook,annapaper}, Karczewska \emph{et al.} produced a single solitary wave solution to the eKdV equation \eqref{eKdV} with the shallow water wave coefficients \eqref{coefs} in the form traveling wave solutions. Specifically, by substituting for $u(x,t) = A_s\mathrm{sech}^2(w_{s}\,(x-V_st))$ into the equation, the authors obtained
\begin{equation} \label{Anna-result1}
   \quad w_s \approx \sqrt{\dfrac{0.145137}{\beta}}, \quad V_s \approx 1.11455, \quad A_s \approx \dfrac{0.242399}{\alpha},
\end{equation}
thereby determining the three unknown wave parameters. A comparison with the results reported reveals several quantitative and qualitative differences. First, the solutions \eqref{Anna-result1} do not appear to reduce (directly) to the classical KdV expressions when the higher order terms in the eKdV equation are omitted, in contrast to the results obtained in the present work and in the study \cite{hamidgk,tim_soliton}. As emphasized earlier, such asymptotic reduction is essential when solitonic solutions are employed in the derivation of DSW solutions. Evidently, the previous expression for $A_s$ captures the qualitative behavior of the amplitude in the presence of resonant effects, although significant quantitative discrepancies remain when compared with direct numerical solutions as presented in Figure \ref{fig:sol_comps}(b). On the other hand, the solitary wave velocity $V_s$ is reported to remain fixed at approximately $V_s \approx 1.11455$, independently of changes in the parameters $\alpha$ and $\beta$. This behavior is valid only for small values of $\alpha$ and $\beta$, which, in fact, has been the standing assumption throughout Sections \ref{sec:3} and \ref{sec:4}, enabling the asymptotic expansion. In this case, Figure \ref{fig:sol_comps}(a) indicates that this wave velocity given provides a reasonable approximation. When these parameters increase—while still satisfying the fundamental assumption $\alpha,\beta< 1$ underlying the derivation of the eKdV equation and its associated solutions—the velocity deviates noticeably from this constant value, as can be seen in Figure \ref{fig:sol_comps}(a). Furthermore, in the context of the application considered in Section \ref{sec:4}, a constant velocity is not well suited when the single solitonic solution is applied to analyze the DSW solutions, where the wave velocity depends monotonically on the magnitude of the initial discontinuity $\Delta$ (and implicitly on $\alpha$ and $\beta$), as presented in Figure \ref{fig:dsw_comps}(a).

A generalization of the results of \cite{annabook,annapaper} was give by Khusnutdinova \emph{et al.} \cite{karimafifth}, where a single solitary wave solution was found for the full eKdV equation \eqref{eKdV}, not limited to the shallow water wave coefficients. Here again, the authors substituted a single solitary wave solution in the form of $\mathrm{sech}^2$ into the eKdV equation and, through careful algebraic manipulations, demonstrated the existence of a \emph{family} of solitons, among which are the so-called \emph{embedded solitons}—isolated localized waves lying within the continuous spectrum of the corresponding linear wave system—and studied their numerical solutions. However, these solutions appear to exist only for a specific range of choices of the coefficients $B_1, B_2, B_3$, and $ B_4$ (see equations (23)--(24) in \cite{karimafifth}). 

The above observations highlight the advantages and practical relevance of the solitonic solutions derived in the present work through the techniques of the calculus of variations. In general, this approach allows us to construct single solitary wave solutions for the eKdV equation \eqref{eKdV} without imposing restrictions on the coefficients of the second-order terms, namely $B_1, B_2, B_3$, and $ B_4$, in contrast to previous works; while at the same time providing explicit second-order formulas for the width, velocity, and height of the solitary wave, making the solution more amenable to applications involving DSWs.

\section{Conclusions}\label{sec:7}

In \cite{hamidgk}, the present authors asked whether the method of averaged Lagrangian, which was used to obtain single solitary wave solutions for the \emph{conservative}-eKdV equation, can be extended to produce a single traveling solitary wave solution to the eKdV equation \eqref{eKdV} for general coefficients $B_1, B_2, B_3$, and $ B_4$. In the present work, we showed that a master or augmented Lagrangian, modeled on the known Luke's Lagrangian, can indeed be constructed, allowing the variational method to determine the complete set of wave parameters for this solitary wave.

Notably, a careful inspection of the derivation of the eKdV equation for shallow water waves \eqref{ekdv-intro} suggested an approach for constructing a Lagrangian for the non-conservative problem that can be used effectively within the averaging method. In this respect, a few key observations were made. First, it was noticed that the Euler-Lagrange equations \eqref{EL-sec2-2} arising from Luke's Lagrangian become compatible with the second-order Boussinesq equations \eqref{Bous2-intro} if the second-order ansatz \eqref{anzats-2-intro} is imposed, together with the first-order Boussinesq equations in the form of equation \eqref{Bous-ansatz}. Moreover, with regard to the latter constraint, it was essential to recognize that all necessary corrections to the Euler–Lagrange equations arise at second-order. The observations were formalized in the traveling wave setting for the general eKdV equation \eqref{eKdV} by incorporating the two constraints via the method of Lagrange multipliers at the proper orders. These, together with Luke's Lagrangian, formed the augmented Lagrangian \eqref{aug-L(theta)} for the derivation of the eKdV equation \eqref{eKdV} for variable coefficients $B_1, B_2, B_3$ and $B_4$\footnote{It is worth noting that, by using constraint \eqref{anzats-2-intro} to express the Luke Lagrangian in terms of a single field $u$ (and its derivatives), the Lagrangian itself becomes model-dependent rather than limited to the derivation of the shallow water wave model, allowing for the formulation of other models in the traveling wave setting.}.

The Euler-Lagrange equations \eqref{EL_av_3} associated with the augmented Lagrangian were then averaged to establish the existence of a single solitary wave with a $\mathrm{sech}^{2}$ profile. In addition, in terms of the  solitonic amplitude parameter $a_0$, explicit formulas were derived for higher order corrections to the wave velocity $V_s$ and the inverse width $w_s$. When combined with basic energy estimates, these results also yielded a second-order expression for the solitary wave height $A_s$. Detailed comparisons with direct numerical simulations of the eKdV solitary waves further demonstrated excellent agreement between theoretical predictions and numerical results. 

An essential feature of these expressions for the velocity $V_s$, inverse width $w_s$ and height $A_s$ is that they reduce to the classical results—namely, those for the KdV equation—when $B_1, B_2, B_3,$ and $B_4$ are all set to zero, a property that distinguishes the present theory from previous works. This asymptotic reduction property renders the theory developed particularly useful for applications in dispersive hydrodynamics. One application considered in this work was the study of DSWs generated by the eKdV Riemann problem. Since the model is non-integrable and exact Whitham modulation equations cannot be derived, solutions for eKdV DSWs are thus approximated using the equal amplitude approximation method introduced by Marchant and Smyth \cite{equalamp}, yielding excellent agreement with numerical results. 

In the traveling single solitary waves setting, variational methods have proven to provide a powerful machinery for constructing solutions to the eKdV equation \eqref{eKdV} without imposing restrictions on the coefficients $B_1$, $B_2$, $B_3,$ and $B_4$, in contrast to previous studies, while at the same time retaining the desirable asymptotic reduction property. A promising direction for future work is to extend these methods to more complex settings, including nonlinear periodic wavetrains, multi-solitonic interactions, and solitary wave propagation over uneven bottom topography.

%%%%%%%%%

\appendix

\section{Augmented Lagrangian derivation of the eKdV model in travelling wave coordinate}\label{appa} 
Here, we present the derivation of the eKdV equation \eqref{eKdV}, in the coordinate $\theta = x-V_st$, from the Lagrangian \eqref{aug-L-hat}. We begin by re-writing the eKdV equation in terms of $\theta$
\begin{equation} \label{a-ekdv}
    -V_s u_\theta + u_\theta + \dfrac{3}{2} \alpha u u_\theta + \dfrac{1}{6} \beta u_{3\theta} + B_1 \alpha^2 u^2 u_\theta + B_2 \beta^2 u_{5 \theta} + \alpha \beta \left( B_3 u_\theta u_{\theta \theta} + B_4 u u_{3\theta} \right)  = 0.
\end{equation}
The Euler-Lagrange system obtained from the augmented Lagrangian
\begin{equation}
L_{\mathrm{aug}} =L_{\mathrm{aug}}(u, u_\theta, u_{2\theta}, u_{3\theta}, u_{4\theta}, \phi_\theta) 
\end{equation}
takes the form 
\begin{comment}
\begin{equation} \label{a-ekdv}
    -V_s u_\theta + u_\theta + \dfrac{3}{2} \alpha u u_\theta + \dfrac{1}{6} \beta u_{3\theta} + B_1 \alpha^2 u^2 u_\theta + B_2 \beta^2 u_{5 \theta} + \alpha \beta \left( B_3 u_\theta u_{\theta \theta} + B_4 u u_{3\theta} \right)  = 0
\end{equation}
The Euler-Lagrange system obtained from the augmented Lagrangian $L_{\mathrm{aug}} =L_{\mathrm{aug}}(u, u_\theta, u_{2\theta}, u_{3\theta}, u_{4\theta}, \phi_\theta) $ is 
\end{comment}
\begin{equation}
  \left\{
  \begin{array}{ll}
  \vspace*{0.1 in}
    \dfrac{\partial L_{\mathrm{aug}}}{\partial u} - \dfrac{d}{d \theta} \left( \dfrac{ 
\partial L_{\mathrm{aug}}}{\partial u_{\theta}} \right) + \dfrac{d^2}{d \theta^2} \left( \dfrac{ 
\partial L_{\mathrm{aug}}}{\partial u_{\theta \theta}} \right)  - \dfrac{d^3}{d \theta^3} \left( \dfrac{ 
\partial L_{\mathrm{aug}}}{\partial u_{3 \theta}} \right) + \dfrac{d^4}{d \theta^4} \left( \dfrac{ 
\partial L_{\mathrm{aug}}}{\partial u_{4 \theta}} \right)=0, \\ \label{a-EL}

\dfrac{d}{d \theta} \left( \dfrac{ 
\partial L_{\mathrm{aug}}}{\partial \phi_{\theta}} \right) = 0.
    \end{array} 
    \right.
\end{equation}
The second equation in system \eqref{a-EL} results in $\lambda_\theta = 0$. On the other hand, the first equation, up to third-order in parameters $\alpha$ and $\beta$, gives
\begin{eqnarray} \label{a-EL2}
    & & \left(-V_s - \lambda \right) + \left(2 + \dfrac{\lambda}{2}  - \dfrac{3V_s}{2}   \right) u \alpha +\left( \dfrac{3 V_s}{4}  - 3\lambda C_1 + \dfrac{3}{4}  -3V_s C_1   \right) u^2 \alpha^2 \nonumber \\
    & & \mbox{} +  \left( (1-V_s) \sigma + \left[ - 2 \lambda C_4  + \dfrac{V_s}{3}  -2V_s C_4  \right] u_{\theta \theta} \right) \alpha \beta + \left( 4C_1  - \dfrac{7}{8} - 4V_sC_1 \right) u^3 \alpha^3 \nonumber \\
   & & + \left( \dfrac{3}{2} u \sigma + \bigg[ \dfrac{-4}{3} + 4C_4 + \dfrac{3V_s}{2} - 4V_sC_4 \bigg] u u_{\theta \theta} + \bigg[ \dfrac{-2}{3} + \dfrac{3V_s}{4} + 2C_4 - 2V_s C_4 \bigg] u_{\theta}^2 
\right) \alpha^2 \beta \nonumber \\
   & & + \left( \dfrac{1}{6} \sigma_{\theta \theta} + \left[  -\dfrac{1}{5} + \dfrac{V_s}{4} -2V_s C_2 + 2C_2   \right] \right) u_{4 \theta} \alpha \beta^2= 0.
\end{eqnarray}
\begin{comment}
\begin{align*} \label{a-EL2}
 (-V_s - \lambda) + \left(2 + \dfrac{\lambda}{2}  - \dfrac{3V_s}{2}   \right) u \alpha +\Bigg( \dfrac{3 V_s}{4}  - 3\lambda C_1 + \dfrac{3}{4}  -3V_s C_1   \Bigg) u^2 \alpha^2 \\
 + \left( (1-V_s) \sigma + \left[ - 2 \lambda C_4  + \dfrac{V_s}{3}  -2V_s C_4  \right] u_{\theta \theta} \right) \alpha \beta  + \left( 4C_1  - \dfrac{7}{8} - 4V_sC_1 \right) u^3 \alpha^3 \\
 + \Bigg( \dfrac{3}{2} u \sigma + \bigg[ \dfrac{-4}{3} + 4C_4 + \dfrac{3V_s}{2} - 4V_sC_4 \bigg] u u_{\theta \theta} + \bigg[ \dfrac{-2}{3} + \dfrac{3V_s}{4} + 2C_4 - 2V_s C_4 \bigg] u_{\theta}^2 
\Bigg) \alpha^2 \beta \\
+ \left( \dfrac{1}{6} \sigma_{\theta \theta} + \left[  -\dfrac{1}{5} + \dfrac{V_s}{4} -2V_s C_2 + 2C_2   \right] \right) u_{4 \theta} \alpha \beta^2= 0 \numberthis
\end{align*}
\end{comment}
The choice
$$\sigma = X u_{\theta \theta}$$
where the parameter $X $ needs to be determined, yields the correct form of the final equation as will be demonstrated below.

To ensure a non-trivial equation of motion at the lowest order, we set $\lambda = -V_s$; that is, the constant of integration for Euler-Lagrange equations \eqref{a-EL}$_2$ is taken to be $-V_s$. Making this identification in equation \eqref{a-EL2}, dividing by $2\alpha$, taking $V_s =1$ at $\mathcal{O}(\alpha, \beta)$ (see equation \eqref{V-intro}), and finally differentiating equation \eqref{a-EL2} with respect to the variable $\theta$ produce
\begin{align} \label{a-EL3}
-V_s u_\theta + u_\theta + \dfrac{3}{2} \alpha u u_\theta + \dfrac{1}{6} \beta u_{3\theta} + \left( 6C_1 -6V_s C_1 - \dfrac{21}{16}  \right) \alpha^2 u^2 u_\theta + \left( C_2 -C_2V_s +\dfrac{V_s}{8} - \dfrac{1}{10} \right) \beta^2 u_{5 \theta} \nonumber \\
 + \Bigg( \bigg[ 4C_4 - 4 V_s C_4 + \dfrac{3 V_s}{2} - \dfrac{4}{3}  + \dfrac{3}{4} X  \bigg] u_\theta u_{\theta \theta} + \bigg[   2C_4 - 2V_sC_4  + \dfrac{3V_s}{4}    - \dfrac{2}{3} + \dfrac{3}{4} X \bigg] u u_{3\theta} 
\Bigg) \alpha \beta = 0.
\end{align}
This is the eKdV equation \eqref{a-ekdv} with  
\begin{equation} \label{a-sys}
  \left\{
  \begin{array}{ll}
  \vspace*{0.1 in}
B_1 = 6C_1 -6V_s C_1 - \frac{21}{16}, \\
  \vspace*{0.1 in}
B_2 = C_2 -C_2V_s +\frac{V_s}{8} - \frac{1}{10}, \\
  \vspace*{0.1 in}
B_3 = 4C_4 - 4 V_s C_4 + \frac{3 V_s}{2} - \frac{4}{3}+ \frac{3}{4} X, \\
  \vspace*{0.1 in}
B_4 =  2C_4 - 2V_sC_4  + \frac{3V_s}{4}    - \frac{2}{3}+ \frac{3}{4} X.
    \end{array} 
    \right.
\end{equation}
Together equations \eqref{a-sys}$_3$ and \eqref{a-sys}$_4$ give $X = \frac{4}{3} (2B_4 - B_3) $. This identification reduces the above system of equations into 
\begin{equation} \label{a-sys-2}
  \left\{
  \begin{array}{ll}
  \vspace*{0.1 in}
B_1 = 6C_1 -6V_s C_1 - \frac{21}{16}, \\
  \vspace*{0.1 in}
B_2 = C_2 -C_2V_s +\frac{V_s}{8} - \frac{1}{10}, \\
  \vspace*{0.1 in}
B_3 - B_4 =  2C_4 - 2V_sC_4  + \frac{3V_s}{4}    - \frac{2}{3}.
    \end{array} 
    \right.
\end{equation}
In practice, the coefficients $B_1, B_2, B_3$ and $B_4$ are known and depend on modeling considerations. Therefore, by solving the system \eqref{a-sys-2}  we can find the corresponding coefficients $C_1, C_2,$ and $C_4$: 

\begin{equation} \label{a-sys-sol}
  \left\{
  \begin{array}{ll}
  \vspace*{0.1 in}
C_1 = \dfrac{16B_1 +21}{96(1-V_s)}, \\
  \vspace*{0.1 in}
C_2 = \dfrac{40B_2 - 5V_s + 4}{40 (1-V_s)}, \\
  \vspace*{0.1 in}
C_4 =  \dfrac{12(B_3 - B_4) -9V_s+8}{24(1 - V_s)}.
    \end{array} 
    \right.
\end{equation}
In particular, in the context of the water wave problem where $B_1= - \frac{3}{8}, B_2 = \frac{19}{360}, B_3 = \frac{23}{24}, B_4 = \frac{5}{12}$, we obtain
\begin{equation} \label{a-sys-sol}
  \left\{
  \begin{array}{ll}
  \vspace*{0.1 in}
C_1 = \dfrac{5}{32(1-V_s)}, \\
  \vspace*{0.1 in}
C_2 = \dfrac{\frac{11}{9}- V_s}{8(1-V_s)},  \\
  \vspace*{0.1 in}
C_4 = \dfrac{\frac{29}{2} - 9V_s}{24(1-V_s)},
    \end{array} 
    \right.
\end{equation}
which will finally produce the eKdV equation (in terms of variable $\theta$) for shallow water waves.

\end{document}